\documentclass[prb,onecolumn,superscriptaddress,longbibliography,aps,11pt,sort,numbers]{revtex4-1}

\usepackage{comment}

\usepackage{graphicx}
\usepackage{bm}
\usepackage{color}
\usepackage{epstopdf}

\usepackage{color,soul}

\usepackage{amsmath}
\usepackage{amssymb}
\usepackage{epstopdf}
\usepackage[raggedright]{titlesec}
\usepackage[normalem]{ulem}
\usepackage{xfrac} 
\usepackage{multirow}

\usepackage[urlcolor=blue,colorlinks=true,citecolor=blue,linkcolor=blue,pdfstart
view={FitH},bookmarks=false]{hyperref}
\usepackage{adjustbox} 
\usepackage[table]{xcolor}

\graphicspath{{fig/}{./fig/}{.}}

\usepackage{tikz,xcolor,hyperref}

\definecolor{lime}{HTML}{A6CE39}
\DeclareRobustCommand{\orcidicon}{%
	\begin{tikzpicture}
	\draw[lime, fill=lime] (0,0)
	circle [radius=0.16]
	node[white] {{\fontfamily{qag}\selectfont \tiny ID}};
	\draw[white, fill=white] (-0.0625,0.095)
	circle [radius=0.007];
	\end{tikzpicture}
	\hspace{-2mm}
}

\foreach \x in {A, ..., Z}{%
	\expandafter\xdef\csname orcid\x\endcsname{\noexpand\href{https://orcid.org/\csname orcidauthor\x\endcsname}{\noexpand\orcidicon}}
}

\renewcommand{\arraystretch}{1.5}
\begin{document}
	\title{Contemporary Insights into Electronic Structure and Microscopic Transport in Nodal-Line Semimetals}


\author{Ashutosh S. Wadge\orcidA}
\email{wadge@magtop.ifpan.edu.pl}
\affiliation{International Research Centre MagTop, Institute of Physics, Polish Academy of Sciences, Aleja Lotnik\'ow 32/46, PL-02668 Warsaw, Poland}

\author{Pardeep K. Tanwar\orcidP}
\email{pardeep$_$kumar.tanwar@fysik.lu.se}

\affiliation{Division of Synchrotron Radiation Research, Lund University, SE-22100 Lund, Sweden}
\affiliation{MAX IV Laboratory, Lund University, SE-22100 Lund, Sweden}
	
\author{Giuseppe Cuono\orcidZ}
\email{giuseppe.cuono@unimib.it}
\affiliation{Department of Materials Science, University of Milan-Bicocca, Via Roberto Cozzi 55, 20125 Milan, Italy}
\affiliation{Consiglio Nazionale delle Ricerche CNR-SPIN, c/o Universit\'{a} degli Studi "G. D’Annunzio", 66100 Chieti, Italy}

\author{Carmine Autieri\orcidO}
\email{autieri@magtop.ifpan.edu.pl}
\affiliation{International Research Centre MagTop, Institute of Physics, Polish Academy of Sciences, Aleja Lotnik\'ow 32/46, PL-02668 Warsaw, Poland}
\affiliation{SPIN-CNR, UOS Salerno, IT-84084 Fisciano (SA), Italy}

	\begin{abstract}
{\bf Topological semimetals have emerged as an important class of quantum materials with novel electronic responses and unconventional transport phenomena. Among them, nodal-line semimetals are distinguished by band crossings that extend along one-dimensional lines in momentum space rather than occurring at discrete points, forming closed loops, chains, or extended lines. The stability of these nodal structures is governed by crystalline symmetries such as mirror, spin-rotation, and nonsymmorphic operations, which give rise to characteristic topological invariants and surface states, including drumhead-like bands.
In this review, we present a comprehensive overview of the theoretical framework and experimental realization of nodal-line semimetals, with particular emphasis on symmetry protection and the consequences of symmetry breaking. We discuss the classification of nodal-line structures, their evolution into other topological phases, and their signatures in electronic structure measurements and transport phenomena. Special attention is given to insights obtained from angle-resolved photoemission spectroscopy and related probes. By bringing together symmetry analysis, band topology, and experimental observations, this review aims to clarify the relationship between topology, magnetism, and measurable electronic responses in nodal-line semimetals. These considerations highlight their potential as a versatile platform for next-generation topological electronic functionalities and emergent quantum phenomena beyond conventional paradigms.
}
	\end{abstract}
	
	\maketitle
	
	\newpage
\section{Introduction}

Topological semimetals have emerged as a central theme in contemporary condensed matter physics owing to their fundamental significance and potential technological applications, ranging from low-power electronics to quantum technologies~\cite{hasan2010colloquium}. Beyond the traditional classification of materials as metals, semiconductors, or insulators, topology has emerged as a unifying principle governing the global structure of electronic wave functions in momentum space~\cite{hasan2010colloquium,hasan2011three}. In this broader landscape, topological semimetals occupy a distinctive position. Unlike topological insulators, where a bulk energy gap separates occupied and unoccupied states, topological semimetals host symmetry-protected band crossings at or near the Fermi level, giving rise to relativistic quasiparticles and unconventional electronic responses~\cite{burkov2016topological,gao2019topological,lv2021experimental}.

Early investigations in this field primarily focused on Dirac and Weyl semimetals, in which conduction and valence bands intersect at isolated points in the Brillouin zone (BZ). These point-like degeneracies act as monopoles of Berry curvature and lead to remarkable phenomena such as Fermi-arc surface states, chiral anomaly–induced negative magnetoresistance, and large anomalous transport coefficients~\cite{armitage2018weyl,hasan2017discovery,yan2017topological,tanwar2022severe}. More recently, however, a broader and conceptually richer class of systems has been identified: nodal-line semimetals (NLSMs) \cite{Fang16}, in which band crossings extend along one-dimensional manifolds in three-dimensional momentum space. In such materials, the degeneracies may form closed loops, extended lines, chains, or interconnected networks, fundamentally modifying the topology and low-energy electronic structure.

The extended nature of nodal lines introduces qualitative differences compared with point-node systems. First, their stability often relies on crystalline symmetries beyond inversion and time-reversal symmetry, including mirror reflection, glide planes, screw rotations, and other nonsymmorphic operations \cite{Fang16,Chang25}. These symmetry constraints can enforce degeneracies along high-symmetry lines or planes and, in certain cases, protect them even in the presence of finite spin--orbit coupling (SOC). Second, the surface projection of nodal loops can host drumhead-like states characterized by enhanced density of states, providing a natural platform for correlation-driven instabilities and surface electronic reconstructions \cite{Chang25}. Third, the torus-shaped Fermi surfaces characteristic of nodal-line systems give rise to distinctive Berry-phase signatures in quantum oscillations and unconventional magnetotransport behavior such as chiral anomaly-induced negative magnetoresistance ~\cite{alexandradinata2023fermiology}.
\begin{figure}[h]
\centering
\includegraphics[width=0.8\linewidth]{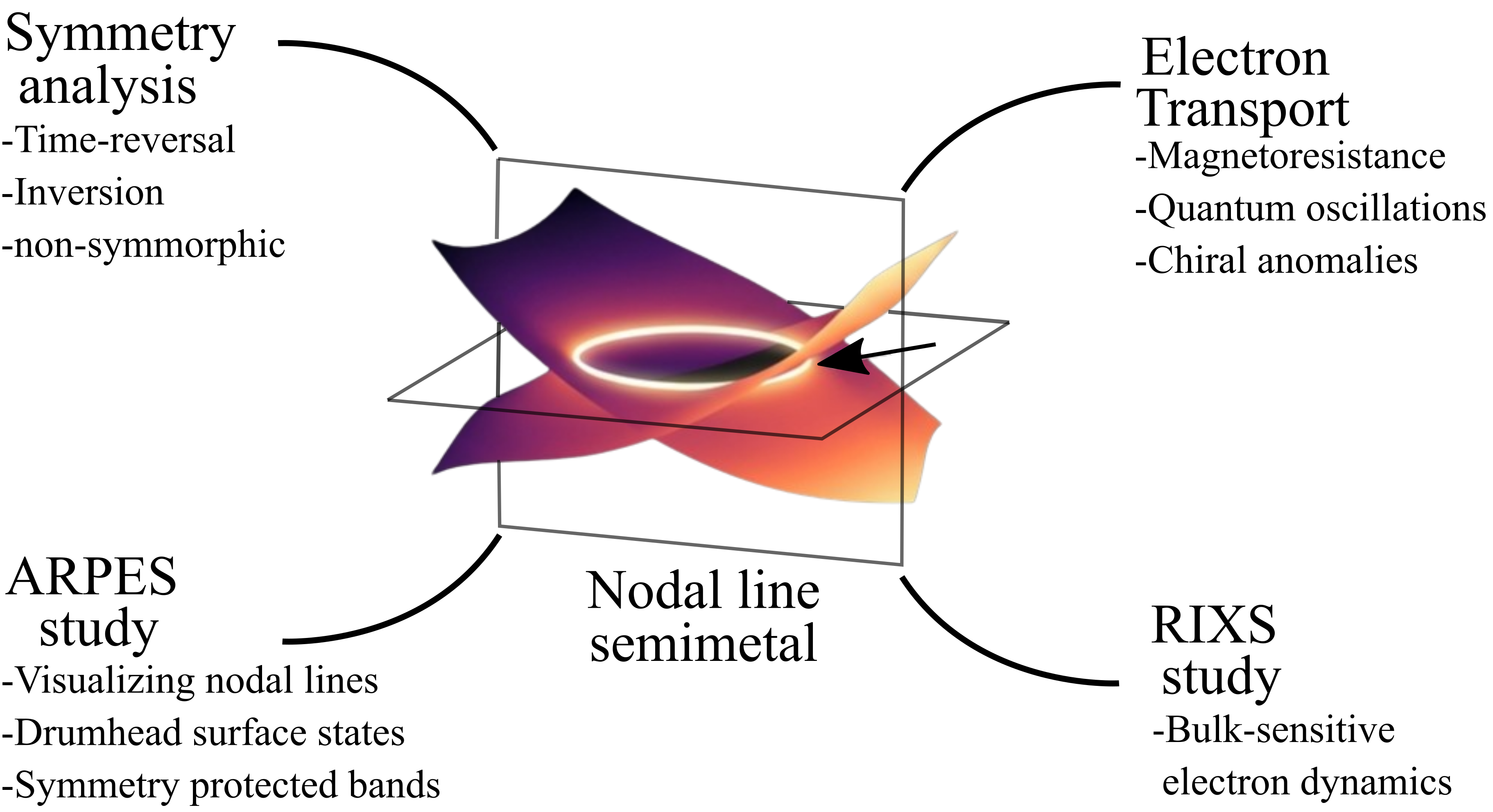}
\caption{Schematic overview of nodal-line semimetals showing a nodal-line band crossing. Symmetry analysis explains its protection, angle-resolved photoemission spectroscopy (ARPES) visualizes the nodal lines and drumhead states, transport measurements probe electronic signatures, and resonant inelastic X-ray scattering (RIXS) provides bulk-sensitive information on elementary excitations.
}
\label{fig1}
\end{figure}

Despite rapid theoretical progress, the unambiguous experimental identification of nodal lines as extended objects in three-dimensional momentum space has remained a significant challenge. Unlike isolated Dirac or Weyl points, which may be inferred from discrete band crossings in a single momentum cut, nodal lines must be traced continuously across the Brillouin zone and distinguished from accidental or surface-derived features. Angle-resolved photoemission spectroscopy (ARPES) has therefore played a central role in establishing the existence of nodal-line topology. By directly resolving the momentum-resolved single-particle spectral function $A(\mathbf{k},\omega)$, ARPES enables visualization of band crossings, systematic photon-energy-dependent mapping of the out-of-plane momentum $k_z$, and polarization-dependent verification of orbital symmetry~\cite{lu2012angle,lv2021experimental}. Through these capabilities, ARPES has played a central role in establishing nodal-line semimetals as experimentally verified quantum materials beyond their initial theoretical predictions.

Experimentally confirmed nodal-line systems now span several materials families, including square-net compounds such as ZrSi$X$ ($X = \mathrm{S, Se, Te}$), transition-metal dipnictides $M$Pn$_2$ ($M = \mathrm{Zr, Hf}$; Pn = $\mathrm{P, As}$), and rare-earth antimonide tellurides $X$SbTe. These platforms provide tunable laboratories in which the robustness, fragility, and controllability of nodal-line topology can be systematically explored. Spin--orbit coupling can partially gap or reshape nodal manifolds, chemical substitution and strain modify symmetry conditions, and magnetic ordering introduces additional symmetry breaking that may either preserve or reconstruct nodal-line degeneracies. The resulting interplay between symmetry, topology, and magnetism gives rise to rich phenomenology observable in transport coefficients, anomalous Hall responses, and quantum oscillations. Recent studies combining ARPES and transport measurements further demonstrate the versatility of these systems as platforms for investigating topological electronic states~\cite{wadge2022electronic}.

Beyond single-particle band topology, nodal-line semimetals also provide promising opportunities to investigate collective excitations and dynamical probes of topology. The presence of extended band crossings modifies particle–hole continua, plasmon modes, and magnetic excitations. Emerging spectroscopic techniques such as resonant inelastic x-ray scattering (RIXS) offer complementary, bulk-sensitive access to electronic dynamics and may enable the reconstruction of symmetry-constrained topological indices beyond the surface sensitivity of conventional ARPES.

In this review, we provide a comprehensive and experimentally grounded overview of nodal-line semimetals (see FIG.~\ref{fig1}), with particular emphasis on symmetry classification, ARPES visualization, magnetic realizations, and macroscopic transport signatures, highlighting their role as a platform for topological electronic responses and phenomena beyond conventional frameworks.
    
\section{Symmetry classification and its consequences on the electronic structures}
	
Topological semimetals are systems in which the conduction and valence bands intersect at points or lines in the BZ. In the absence of symmetry constraints, electronic bands approaching each other in energy typically hybridize and open a finite gap due to level repulsion. In contrast, crystalline or time-reversal symmetries can protect these band crossings, as the bands carry different quantum numbers. Topological semimetals represent, therefore, symmetry-protected topological phases of matter. When the bands cross at specific points, we have Dirac or Weyl semimetals \cite{yan2017topological}, while when the bands cross along curves in the BZ, we have the nodal-line semimetals (NLSM) \cite{Fang16}. A minimal model capturing the essential features of a nodal-line semimetal can be formulated as follows \cite{PhysRevX.11.031017}: 
\begin{equation}
H(\mathbf{k}) =
\left(6t - t_1 - 2t \sum_{i=x,y,z} \cos k_i \right)\sigma_z + 2t_2 \sin k_z  \sigma_x .
\end{equation}
Here $\sigma_x$ and $\sigma_z$ are the Pauli matrices. The parameters $t$, $t_1$, and $t_2$ determine the size of the nodal line and the Fermi velocity.
Topological semimetals with both nodal points and nodal lines can also be observed \cite{Luo25coexisting}. Nodal lines can be classified according to the symmetries that protect them. 
Unlike topological insulators, in topological semimetals, the conduction and valence bands intersect at certain points, preventing a clear distinction between occupied and unoccupied bands and making the definition of a topological invariant less straightforward.
The methodology used to define the topological invariant in these cases is to enclose the node within an imaginary manifold that does not cross it. Since the conduction and valence bands do not cross each other inside the manifold, the previous definition of the topological invariant can be used \cite{Fang16}. 
Mirror-protected nodal lines occur when two non-degenerate bands with opposite mirror eigenvalues cross within a mirror-invariant plane. Because Bloch states in the mirror plane can be labeled by their mirror eigenvalues, bands belonging to different mirror sectors cannot hybridize. As a result, their crossings are symmetry-protected. 

Referring to the definition of topological invariants in the nodal lines discussed above, in this case, the topology can be characterized by an integer $\mathbb{Z}$ invariant. This invariant is defined as the difference in the number of occupied bands belonging to the two mirror symmetry sectors \cite{Fang16}. More specifically, one considers a zero-dimensional enclosing manifold by selecting two points, $p_1$ and $p_2$, located on opposite sides of the nodal line. At these points, the valence and conduction bands are energetically separated, allowing for an unambiguous identification of the occupied states. One can then count the number of occupied bands, $N_1$ and $N_2$, that carry a given mirror eigenvalue at $p_1$ and $p_2$, respectively. The topological invariant is defined as the difference $z_0 = N_1 - N_2$. A nonzero value of this invariant signals a topologically protected band crossing. In particular, when $z_0 = 1$, the crossing occurs between two bands with opposite mirror eigenvalues and is therefore protected by mirror symmetry.
Although nodal-line semimetal materials can also exist in two-dimensional materials~\cite{Feng17,doi:10.1021/acsnano.3c12753,tian2026kagomegoldeneflatbands} and several models have been proposed~\cite {PhysRevB.97.064513,Fernandez24, wu2023hybrid}, the experimental confirmations are comparatively rare. In two-dimensional band structures, accidental band crossings generically occur at isolated points, as hybridization between bands typically opens a gap when they approach each other in energy. Stabilizing an extended line of degeneracy, therefore, requires additional symmetry constraints that prevent band mixing over a continuous region of momentum space. As a result, nodal lines in two-dimensional systems typically rely on specific crystalline symmetries such as mirror or nonsymmorphic operations. This review focuses on nodal lines in three-dimensional materials, since they are more popular and generally more stable against SOC.
Some representative systems in which nodal lines protected by mirror symmetry have been proposed theoretically and, in some cases, observed experimentally include PbTaSe$_2$ \cite{Bian16a}, PbTaS$_2$ \cite{Sun17}, TlTaSe$_2$ \cite{Bian16b}, and the two-dimensional Cu$_2$Si \cite{Feng17}. In materials such as TaAs \cite{Weng15}, or CaAgX (X = P, As) \cite{Yamakage16,PhysRevX.11.031017}, nodal lines were predicted only in the absence of spin-orbit coupling and are generally gapped or transformed into Weyl points once SOC is included. Mirror symmetry-protected nodal lines can be engineered in order to have flat-band features at the Fermi level. A notable example is provided by short-period HgTe/CdTe or HgTe/HgSe superlattices, where \textit{ab initio} calculations have shown that interface-induced symmetry breaking and strain engineering give rise to pairs of circular nodal lines lying in planes parallel to the $k_x$–$k_y$ plane and located at finite $k_z$ \cite{Islam22}. These nodal lines can be understood as remnants of the mirror symmetry-protected nodal lines of the bulk HgTe, which survive despite the breaking of almost all mirror symmetries.
In systems with inversion, time-reversal, and SU(2) spin-rotation symmetries (i.e., in the absence of SOC), nodal lines arise from crossings of non-degenerate bands and are described by two $\mathbb{Z}_2$ invariants: one indicating the existence of the line, and the other its global stability, such that lines with nontrivial invariants can only be annihilated in pairs \cite{Fang16}. Representative proposed examples of this category are Ca$_3$P$_2$ \cite{Xie15}, CaP$_3$ \cite{Xu17} and Cu$_3$(Pd,Zn)N \cite{Kim15}. Double-nodal lines were predicted to appear in spinful systems~\cite{Fang15,Chen2015,PhysRevB.97.081105} with spin-orbit coupling when both conduction and valence bands are doubly degenerate and are protected by a twofold screw rotation in addition to inversion and time-reversal. These fourfold crossings are characterized by an integer $\mathbb{Z}$ invariant. 
These kinds of nodal lines were proposed in SrIrO$_3$ \cite{Fang15,Chen2015,PhysRevB.97.081105} and in the quasi-one-dimensional family BaMX$_3$
(M = V, Nb, or Ta; X = S or Se) \cite{Liang16}.
Depending on the symmetry that protects the nodal line, nodal-line semimetals can be classified into three categories\cite{Fang16,Yang18}: Type A, protected by mirror symmetry; Type B, protected by the combination of inversion and time-reversal symmetries; and Type C, corresponding to double-nodal lines.
Nodal-line phases protected by time-reversal symmetry in achiral, non-centrosymmetric materials are referred to as Kramers nodal lines, as they exhibit Kramers degeneracy \cite{Xie2021}. This class of materials has been investigated in several experimental studies \cite{ld4p-hl13,Sarkar2023} as well as in theoretical works on topological Fermi arcs \cite{5f8t-8qhk}.
Nodal-line semimetals protected by nonsymmorphic symmetries show crossings of the bands on a glide mirror plane or screw rotation axis.
Interesting cases are those of nodal lines protected by nonsymmorphic symmetries and in the presence of time-reversal symmetry, where along any path on the glide-invariant plane connecting two distinct time-reversal invariant momenta, the bands are forced to exhibit an hourglass-like connectivity, which leads to a band crossing along the path \cite{Chang25}. The band crossings form a nodal-ring structure on the nonsymmorphic-invariant plane.
Other systems showing nodal lines protected by nonsymmorphic symmetries are: ZrSiS \cite{schoop2016dirac,Neupane16}, HfSiS \cite{Takane16}, HfP$_2$ \cite{sims2020termination}, ZrP$_2$ \cite{bannies2021extremely}, ZrAs$_2$ \cite{Zhou21,Wadge21} and Zr$_5$Pt$_3$\cite{Bhattacharyya2021-qy}.
In the presence of nonsymmorphic symmetries, hourglass nodal lines can be realized. Hourglass fermions are surface fermions whose dispersion is shaped like an hourglass, since the combination of time-reversal symmetry and the nonsymmorphic symmetry enforces a switching of Kramers partners along high-symmetry lines, creating a protected crossing that cannot be removed without breaking the symmetry \cite{Wang16Hourglass,Wang19Hourglass2D,Wang17Hourglass3D}. It was shown that in rhenium dioxide, an hourglass Dirac chain is generated \cite{Wang17Hourglass}, where the neck crossing point of the hourglass forms a Dirac chain, a series of connected four-fold-degenerate Dirac loops in momentum space. An hourglass Dirac loop has also been predicted in the X$_3$SiTe$_6$ (X = Ta, Nb) family \cite{Li18XSiTe}, while coexistence of nodal lines and hourglass has been found experimentally in Nb$_3$SiTe$_6$ \cite{Liu22Ni3SiTe6}. Furthermore, magnetic hourglass fermions were predicted in various systems \cite{Hu22MagneticHourglass,Fakhredine23}. Systems hosting hourglass fermions usually also exhibit semi-Dirac points at the border of the Brillouin zone, where the bands display linear dispersion in one direction and parabolic dispersion in the other\cite{PhysRevMaterials.3.095004}.
Nodal lines can also be classified based on the shape of their band dispersions. Type-I corresponds to the conventional case, where the two crossing bands have opposite slopes around the loop, forming a cone-like crossing. Type-II occurs when the two bands have slopes of the same sign along a particular transverse direction, meaning they are both electron-like or both hole-like, resulting in a strongly tilted linear spectrum. Type-III represents the intermediate case, in which one of the bands is nearly flat along a certain direction. Examples of Type II are K$_4$P$_3$ \cite{Li17} and Mg$_3$Bi$_2$ \cite{Chang19}, while a Type III candidate is the Kagome superconductor CsTi$_3$Bi$_5$ \cite{Yang23}.
It is also possible to have nodal lines represented by quadratic or cubic Hamiltonians, enforced and protected by multiple symmetries \cite{Yu19quadratic}. Quadratic and cubic nodal lines can arise when multiple crystalline symmetries forbid linear terms in the effective $k\!\cdot\!p$ Hamiltonian, thereby stabilizing higher-order dispersions.
The interplay of nodal line topology and magnetism or superconductivity has recently emerged as a central topic in condensed matter physics. In nodal-line semimetals, the presence of nearly flat drumhead surface states enhances correlation effects, which are strong due to the enhancement of the density of states \cite{Chang25}. From a theoretical point of view, this situation promotes Stoner-type ferromagnetic order on the surface, which can open a gap in the surface states and lead to the emergence of a surface Chern insulating phase. In the bulk, although the density of states vanishes when the nodal lines are exactly at the Fermi level and no other bands are present, strong interactions can still induce magnetic ordering, as antiferromagnetic or ferromagnetic orderings depending on the dominant interaction channel. These theoretical considerations have been confirmed by experimental studies. Combined experimental and theoretical studies have demonstrated that the antiferromagnet YMn$_2$Ge$_2$ hosts a Dirac nodal line at the boundary of the Brillouin zone \cite{yang2024topological}. This nodal line is enforced by the interplay of magnetic order, space-time inversion symmetry, and nonsymmorphic crystal symmetries, and involves Mn $d$-orbitals, thus realizing an antiferromagnetic Hund nodal line \cite{yang2024topological}.
Another family that emerged as a fertile playground for exploring the interaction of electronic correlations and magnetic ordering with the nodal line band topology is that of XSbTe (X=Pr, Nd, Gd, Tb, Dy, Er) compounds \cite{Regmi23,Regmi24,Elius25,elius2025electronic,valadez2025low}. In all the compounds of the family, experimental investigations were supported by ab-initio analysis showing the presence of Dirac crossings that correspond to a nodal line along some lines of the Brillouin zone. An eightfold degenerate Dirac nodal line has been predicted in the antiferromagnet Mn$_5$Si$_3$ without SOC, which becomes a fourfold nodal line protected by the combination of time-reversal, partial translation, and nonsymmorphic symmetries when SOC is introduced \cite{Mendoza25}. Instead, in MnC$_4$, a double degenerate Dirac-nodal line has been identified in the absence of SOC \cite{Fernandez24}.
It has been shown that the direction of magnetization can control the topological band features of magnetic materials. For example, in Co$_3$Sn$_2$S$_2$, for a specific in-plane magnetization direction, a gapless nodal loop emerges \cite{Schilberth25}.  
In two-dimensional systems, both ferromagnetic and antiferromagnetic topological nodal-line semimetals have been predicted. Antiferromagnetic candidates include MnC$_4$ \cite{Fernandez24}, MoB$_3$ \cite{Gao24MoB3}, CrB and FeB monolayers \cite{Yan25TMB}, as well as van der Waals heterostructures such as germanene/Mn$_2$S$_2$ \cite{Lv22}.
Recently, nodal line topology has also been predicted in the emerging class of altermagnets, where the interplay between crystal symmetries and unconventional spin splitting can generate Weyl nodal lines or nodal loops in the electronic structure \cite{Antonenko25,Qu25Altermagnetic}.

\section{Angle-Resolved Photoemission Spectroscopy of Nodal-Line Semimetals}
\begin{figure}[h]
\centering
\includegraphics[width=1.0\linewidth]{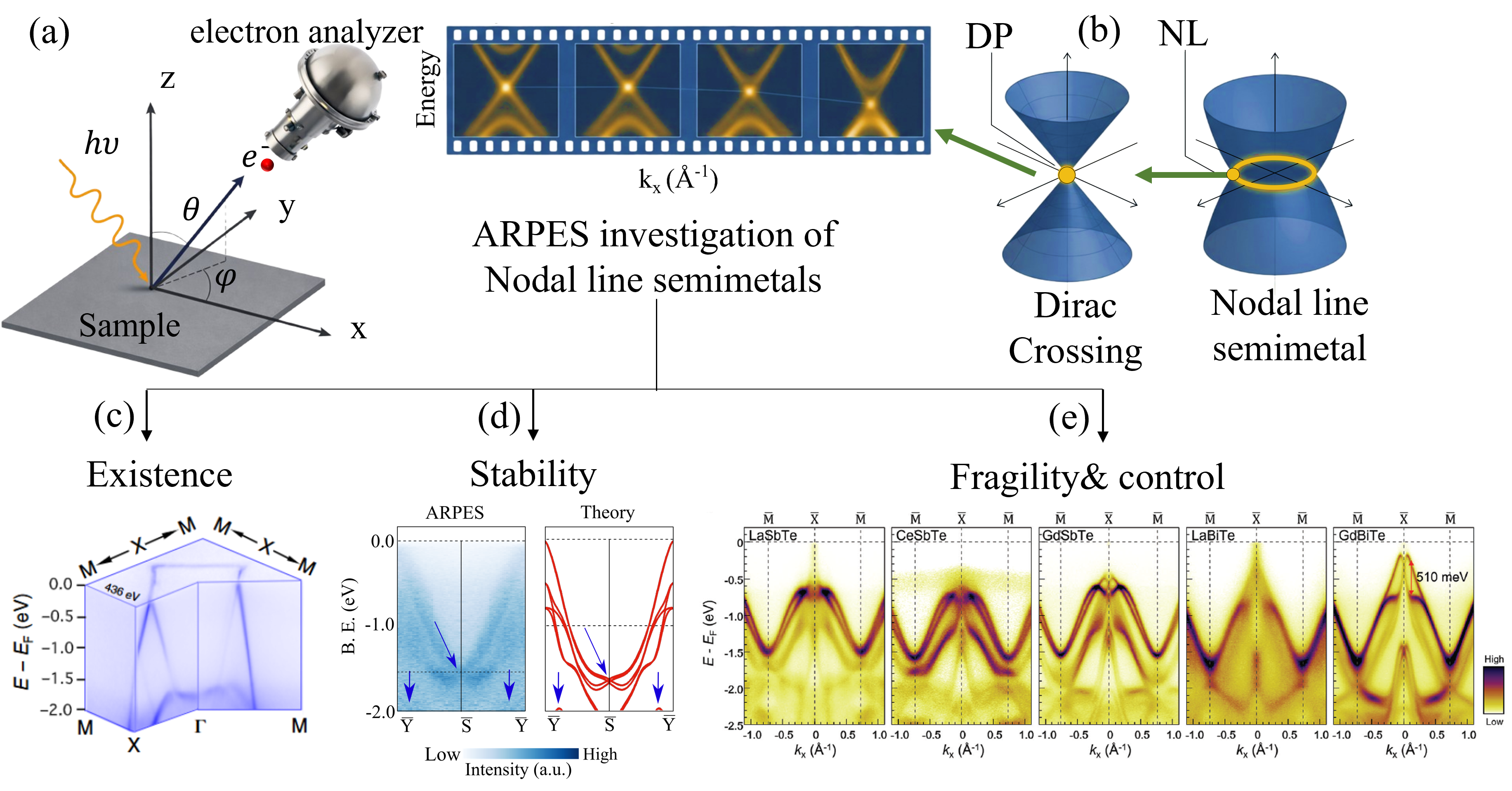}
\caption{ARPES investigation of nodal-line semimetals. (a) Schematic illustration of the ARPES setup. Incident photons ($h\nu$) excite photoelectrons from the sample surface, and the kinetic energy and emission angles ($\theta$, $\phi$) are analyzed to reconstruct the electronic band dispersion in momentum space.
(b) Comparison between a Dirac point (DP), characterized by a linear band crossing at a discrete point, and a nodal-line (NL) semimetal, where the band crossing forms a continuous closed loop in momentum space.
(c) Calculated three-dimensional bulk band structure in the first Brillouin zone showing linear band crossings forming a nodal loop. Energies are referenced to $E_F$. 
(d) ARPES data compared with theoretical calculations, evidencing symmetry-protected degeneracy and robustness against spin--orbit coupling in ZrAs$_{2}$. 
(e) ARPES dispersions of LaSbTe, CeSbTe, GdSbTe, LaBiTe, and GdBiTe demonstrating SOC-induced gap opening and tunability via strain and doping. 
Adapted from references.~\cite{bannies2026electronic, fu2019dirac, Wadge21, cai2025observation}
}
\label{fig2}
\end{figure} 
\subsection{Experimental capabilities and limitations of ARPES}

ARPES provides direct access to the momentum-resolved single-particle spectral function $A(\mathbf{k},\omega)$. It is therefore uniquely suited to investigate the band topology of nodal-line semimetals \cite{damascelli2004probing, yan2017topological, schoop2018chemical}. In contrast to transport measurements, which probe integrated responses of the Fermi surface, ARPES resolves the detailed dispersion of electronic states in energy--momentum space and can directly visualize band crossings that define nodal-line fermions\cite{yang2018visualizing, lv2019angle}.
For three-dimensional nodal-line systems, however, ARPES analysis requires careful consideration of several experimental constraints. While the in-plane momentum components $k_x$ and $k_y$ are determined directly from emission angles, reconstruction of the out-of-plane momentum $k_z$ relies on photon-energy variation and assumptions regarding the inner potential~\cite{iwasawa2020high, kumar2020topological, lv2021experimental, sobota2021angle}.
Systematic photon-energy--dependent measurements are therefore essential to establish whether an observed crossing is truly three-dimensional or merely a two-dimensional surface-related feature\cite{damascelli2004probing}.
Surface sensitivity presents another nontrivial complication. Most vacuum-ultraviolet ARPES experiments probe electronic states within only a few atomic layers of the surface. In materials where surface reconstructions or terminations modify the near-surface electronic structure, surface states may hybridize with bulk states or introduce additional dispersive features. Distinguishing bulk nodal lines from surface resonances requires consistency across photon energies and, where possible, comparison with soft x-ray ARPES measurements that enhance bulk sensitivity \cite{lv2021experimental,sobota2021angle}.
Matrix-element effects further influence the visibility of specific orbitals in ARPES spectra. Since nodal-line crossings often involve bands of distinct orbital character, polarization-dependent measurements are frequently necessary to verify the symmetry origin of the crossing. The absence of spectral weight in a particular geometry does not necessarily imply a gap, but may instead reflect photoemission selection rules\cite{schusser2024towards, figgemeier2025imaging}. 
Within the context of nodal-line semimetals, ARPES provides a direct experimental framework for addressing several fundamental questions. At the most basic level lies the question of existence: whether nodal lines manifest in real materials as extended band degeneracies in momentum space rather than isolated crossing points. Closely related is the issue of stability under what symmetry conditions these degeneracies persist in the presence of realistic perturbations, such as finite spin--orbit coupling or magnetic order? Equally important is the problem of fragility, which mechanisms lift the degeneracy and open measurable gaps along the nodal manifold. Finally, a forward-looking question concerns control: whether nodal-line dispersions can be tuned predictably and systematically through chemical substitution, structural modification, or magnetic interactions (FIG. \ref{fig2}). In the following sections, nodal-line systems are evaluated against these criteria.

\subsubsection{Experimental establishment of extended nodal lines}
\label{subsec:existence_nodal_lines}

The most direct experimental test of nodal-line topology is whether ARPES can resolve band crossings that persist as continuous objects in momentum space rather than appearing as isolated Dirac points. A clear benchmark is provided by the square-net semimetal ZrSiS, where successive momentum cuts
reveal linear crossings that extend over large portions of the Brillouin zone, forming a characteristic diamond-shaped Fermi surface associated with Dirac line nodes~\cite{schoop2016dirac,Neupane16,hosen2017tunability}.
The connectivity of these dispersions across adjacent momentum cuts demonstrates that the degeneracies form one-dimensional manifolds in momentum space rather than discrete Dirac points, establishing ZrSiS as a reference system for ARPES-based identification of nodal-line states.

Closely related compounds such as ZrSiTe exhibit similar momentum-space connectivity~\cite{topp2016non,muechler2020modular}. Although relativistic effects modify portions of the dispersion, ARPES measurements still resolve band crossings that can be traced continuously across the Brillouin zone, reflecting their nodal-line origin.

Beyond square-net systems, recent ARPES studies on ZrAs$_2$ further confirm the existence of multiple nodal-line crossings in a nonsymmorphic crystal structure~\cite{Wadge21}. Band crossings observed along various directions in the Brillouin zone exhibit clear momentum continuity and agree with bulk calculations, supporting their identification as segments of symmetry-enforced nodal lines. In addition, surface-projected features associated with van Hove singularities have been resolved by ARPES in ZrAs$_2$, highlighting the coexistence of bulk nodal-line topology and surface electronic instabilities~\cite{hossain2025superconductivity}.

Recent ARPES investigations of non-centrosymmetric rhombohedral transition-metal dichalcogenides have further illustrated the importance of momentum-selective and bulk-sensitive measurements for resolving topological nodal structures. In particular, micro-focused ARPES combined with first-principles calculations has revealed Kramers nodal lines crossing the Fermi level in 3R-TaS$_2$ and 3R-NbS$_2$, where spin–orbit coupling and crystalline symmetries stabilize exotic open octdong and spindle-torus Fermi surfaces~\cite{Domaine2025}. The study demonstrated that limited out-of-plane momentum resolution in vacuum-ultraviolet ARPES can broaden the apparent $k_z$ dependence of the nodal crossings, while polarization- and momentum-dependent measurements remain essential for distinguishing symmetry-enforced bulk degeneracies from surface-derived features. These results further highlight the growing capability of modern ARPES methodologies to resolve complex three-dimensional nodal manifolds experimentally in layered quantum materials.

Taken together, these experiments establish that nodal lines are experimentally resolvable as extended band degeneracies in momentum space rather than isolated crossing points.
\subsubsection{Experimental validation of nodal-line stability}
\label{subsec:stability}

Having established the existence of extended nodal lines, we now examine their experimental stability under realistic conditions. In this section, we do not revisit the theoretical symmetry analysis presented earlier, but instead focus on whether ARPES measurements confirm the persistence of nodal-line degeneracies in real materials, including in the presence of finite spin--orbit coupling and magnetic order.

In square-net systems such as ZrSiS, ARPES measurements reveal that the linear band crossings predicted by nonsymmorphic symmetry appear precisely along the expected high-symmetry directions and remain continuous across
adjacent momentum cuts~\cite{schoop2016dirac,Neupane16,hosen2017tunability}. 
The persistence of these crossings over extended regions of momentum space demonstrates that the degeneracies are not accidental features of the band structure, but are experimentally robust manifestations of the underlying crystalline symmetry. A similar experimental validation is observed in ZrAs$_2$. As discussed in Section~\ref{subsec:existence_nodal_lines}, ARPES measurements resolve multiple symmetry-enforced nodal lines (NL1--NL5) distributed throughout the Brillouin zone~\cite{Wadge21}. The observed crossings occur at momentum locations consistent with theoretical predictions and exhibit clear momentum continuity, confirming that the nodal-line network remains intact under realistic experimental conditions.

The stability of nodal lines becomes particularly nontrivial in magnetic materials, where time-reversal symmetry is broken. In YMn$_2$Ge$_2$, soft x-ray ARPES measurements reveal nodal-line-derived band crossings that persist in the antiferromagnetically ordered phase, indicating that the band degeneracies survive in the presence of staggered magnetic exchange fields~\cite{yang2024topological}. Likewise, ARPES investigations of the rare-earth antimonide telluride series $R$SbTe demonstrate that nodal-line crossings remain observable
across magnetic ordering temperatures in lighter rare-earth members of the family~\cite{Regmi23,Regmi24,yuan2024observation,hosen2018discovery}.

These experimental observations confirm that nodal-line states can remain stable even in magnetically ordered systems, provided that the relevant crystalline or magnetic symmetry operations are preserved.

\subsubsection{Spin--orbit coupling as a source of fragility and control}
\label{subsec:soc_fragility_control}

Spin--orbit coupling plays a central role in determining both the robustness and the tunability of nodal-line semimetals. ARPES measurements directly reveal both the fragility and tunability associated with relativistic effects.

A prototypical example of SOC-driven evolution is provided by the
square-net series ZrSiX (X = S, Se, Te). In ZrSiS, ARPES measurements reveal nodal-line crossings that remain effectively gapless within typical experimental resolution (approximately 10--20 meV)~\cite{schoop2016dirac,Neupane16,hosen2017tunability, PhysRevB.100.205124}. Upon substitution with heavier chalcogens, measurable lifting of degeneracy emerges. In ZrSiTe, ARPES spectra resolve gap openings along segments of the nodal loop with magnitudes on the order of tens of meV depending on momentum location~\cite{hosen2017tunability,muechler2020modular}. The systematic evolution across ZrSiS $\rightarrow$ ZrSiSe $\rightarrow$ ZrSiTe thus provides a controlled platform for visualizing SOC-induced modification of nodal-line topology.

A similar relativistic tuning mechanism operates in the rare-earth
antimonide telluride family $R$SbTe. ARPES measurements reveal composition-dependent modification of nodal-line-derived crossings: lighter rare-earth members exhibit crossings that appear nearly gapless within experimental resolution, whereas heavier members show finite SOC-induced gaps~\cite{Regmi23,Elius25,valadez2025low,elius2025electronic}. Reported gap magnitudes range from approximately 10 meV up to 60 meV, depending on composition and momentum position.

The pnictide family $M\mathrm{Pn}_2$ ($M$ = Zr, Hf; $\mathrm{Pn}$ = As, P) provides an additional materials platform for systematically varying relativistic effects. In ZrAs$_2$, ARPES measurements resolve multiple symmetry-enforced nodal lines that remain robust within experimental resolution~\cite{Wadge21}. In contrast, HfP$_2$ exhibits partial SOC-induced gap opening along portions of the nodal loop~\cite{sims2020termination}, while ZrP$_2$ shows nodal-loop-derived Fermi surfaces with small but finite splittings consistent with relativistic effects~\cite{bannies2021extremely}. ARPES measurements on HfAs$_2$ likewise reveal nodal-line features modified by SOC and lattice effects~\cite{muhammad2024electronic}.

Taken together, ARPES measurements across the ZrSiX, $R$SbTe, and
$M\mathrm{Pn}_2$ families demonstrate that spin--orbit coupling
constitutes both a fundamental limitation and a powerful tuning
parameter for nodal-line semimetals (see FIG.~\ref{fig2}).

\subsection{ARPES studies of the ZrSiX (X = S, Se, Te) family}
\begin{figure}[h]
\centering
\includegraphics[width=0.8\linewidth]{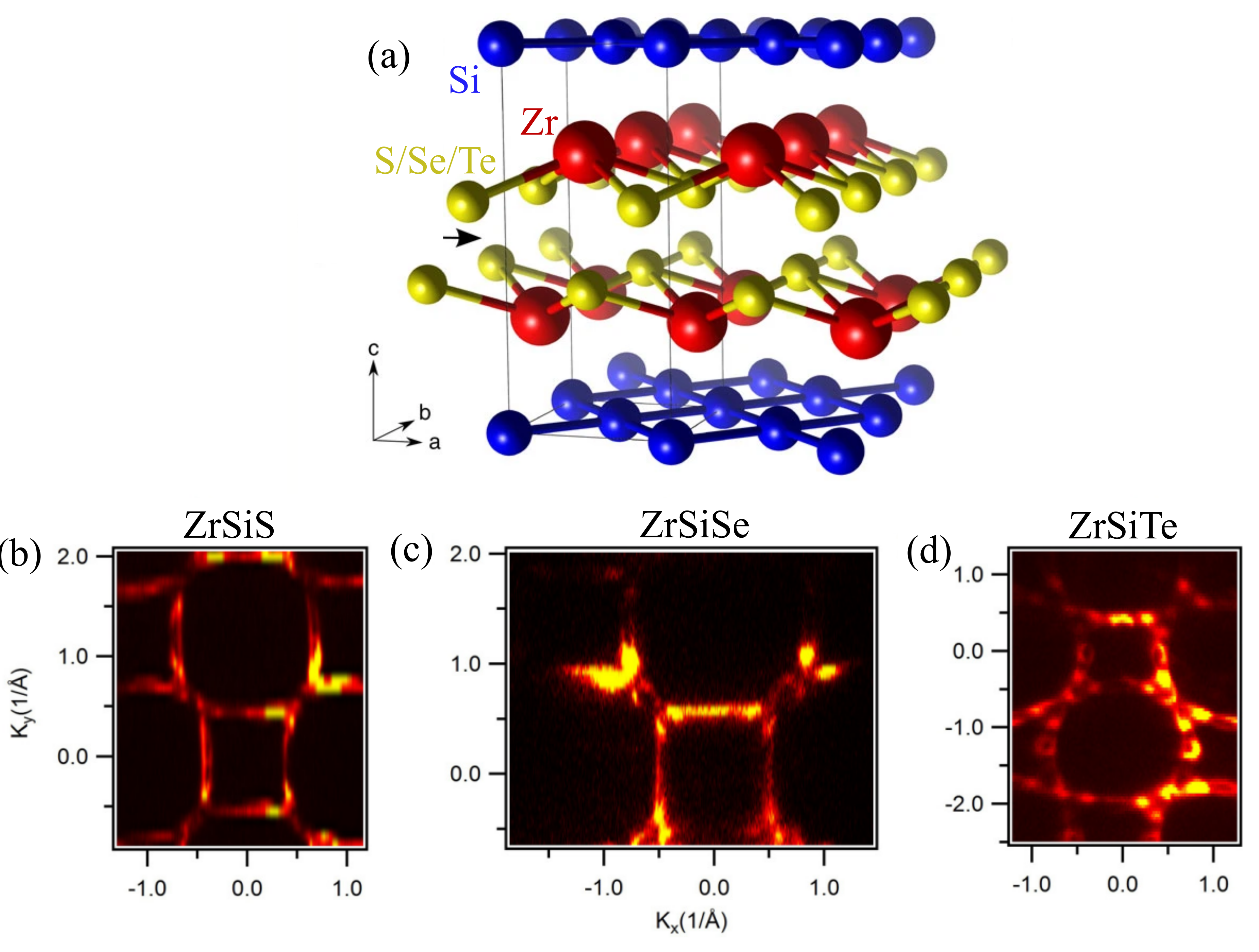}
\caption{ (a) Crystal structure of ZrSi$X$ ($X$ = S, Se, Te), highlighting the Si square-net layer responsible for the Dirac nodal-line states. (b–d) ARPES Fermi-surface maps of ZrSiS, ZrSiSe, and ZrSiTe, showing the characteristic diamond-shaped Fermi contours in the $k_x$–$k_y$ plane and their evolution with increasing spin–orbit coupling. Adapted from references.~\cite{schoop2016dirac,hosen2017tunability}
}
\label{fig3}
\end{figure}
The ZrSiX family constitutes the prototypical square-net
nodal-line platform in which extended Dirac crossings
were experimentally established and heavily studied by ARPES.
Initial measurements on ZrSiS revealed multiple linearly
dispersing bands forming a diamond-shaped Fermi
surface centered at the $\bar{\Gamma}$ point
\cite{fu2019dirac, schoop2016dirac, Neupane16}.
Successive momentum cuts
demonstrated that these crossings are not isolated
Dirac points but instead belong to a continuous nodal-line manifold extending along high-symmetry directions of the Brillouin zone.
\cite{Neupane16}.

ARPES measurements resolved linear dispersions
persisting over an unusually large energy window,
approaching $\sim$2 eV in some momentum regions,
with a purely linear regime of approximately 0.5 eV
around the Fermi level \cite{schoop2016dirac}.
The extracted Fermi velocity was found to be
$v_F \approx 4.3$ eV\AA, comparable to graphene,
highlighting the exceptionally steep character
of the Dirac-like bands.
Unlike atomic Dirac semimetals such as Cd$_3$As$_2$, the crossings in ZrSiS originate from square-net bonding and
band folding induced by nonsymmorphic symmetry,
leading naturally to a nodal-line topology
\cite{schoop2016dirac,topp2017surface}.

Importantly, spin--orbit coupling does not eliminate the nodal-line topology in ZrSiS; instead, it introduces small gaps ($\sim 10$--$20$~meV) at certain crossings, while symmetry-protected degeneracies remain nearly intact. \cite{schoop2016dirac}.
Within typical ARPES resolution (10--20 meV), the crossings appear nearly gapless,
thereby realizing an almost ideal nodal line
semimetal in experiment.

Systematic ARPES comparison across the ZrSiX series reveals the tunability of the nodal line
phase through increasing SOC strength, as shown in FIG.~\ref{fig3}
\cite{hosen2017tunability}.
While ZrSiSe shows a
band structure closely resembling that of ZrSiS,
ZrSiTe exhibits clear modifications near the
$\bar{M}-\bar{\Gamma}-\bar{M}$ direction. In ZrSiTe, ARPES
detects additional Fermi-surface features and theoretical calculations indicate the opening
of a gap of approximately 60 meV along portions
of the nodal-line direction, signaling a
transition from an ideal nodal-line semimetal
toward a nodeless gapped phase as the chalcogen
atomic number increases \cite{hosen2017tunability}.

Beyond bulk nodal-line features, ARPES has
also resolved intense surface-derived states
near the $\bar{X}$ point \cite{topp2017surface}.
These states originate from the lifting of
nonsymmorphic symmetry constraints at the
surface, giving rise to so-called “floating”
two-dimensional bands. Photon-energy-dependent
measurements confirm their surface character
through the absence of $k_z$ dispersion
\cite{topp2017surface}.
Although distinct from the topological drumhead
states, these surface bands coexist with the
bulk nodal-line dispersions and must be
carefully disentangled in ARPES analysis. More recently, high-resolution ARPES measurements revealed the existence of Dirac nodal surfaces in ZrSiS in addition to the previously identified
nodal lines, further enriching the topological band structure of the square-net platform
\cite{fu2019dirac}.

Taken together, ARPES investigations of the
ZrSiX family established: (i) the experimental realization of extended
nodal-line manifolds, (ii) the unusually large linear dispersion
energy scale derived from square-net bonding, and (iii) the systematic evolution of the nodal-line topology under controlled SOC enhancement. These materials, therefore, serve as a benchmark platform for understanding
nodal-line physics in crystalline solids.


\subsection{Transition metal dipnictide family MPn$_{2}$ (M = Zr, Hf; Pn = As, P)} 
\begin{figure}[h]
\centering
\includegraphics[width=0.9\linewidth]{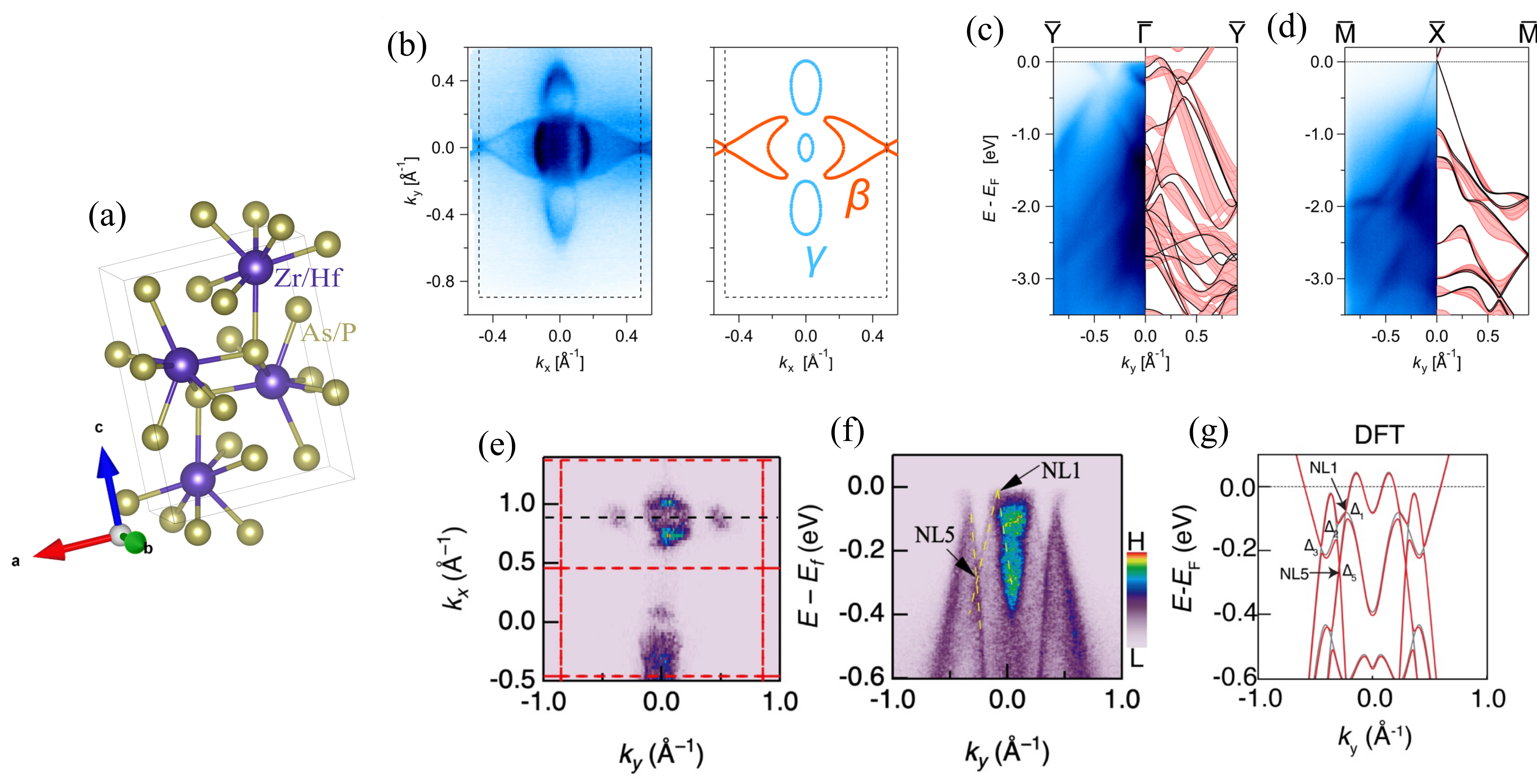}
\caption{
Comparison of bulk nodal-line electronic structures in Zr$X_2$ compounds.
(a) Crystal structure of Zr$X_2$ ($X$ = P, As) in the nonsymmorphic Pnma structure. (b) ARPES Fermi surface of ZrP$_2$ at $k_z \approx 0$ with corresponding DFT calculation. (c,d) Bulk band dispersions of ZrP$_2$ along high-symmetry directions, showing agreement between ARPES and projected DFT bands. (e) Fermi surface of ZrAs$_2$ measured by ARPES. (f) ARPES dispersion along $\bar{Y}-\bar{\Gamma}-\bar{Y}$ revealing nodal-line
crossings (NL1 and NL5). (g) Corresponding DFT band structure including spin–orbit coupling, indicating small SOC-induced gaps at the nodal lines.
Panels (b–d) adapted from reference \cite{bannies2021extremely, hossain2025superconductivity}
}
\label{fig4}
\end{figure}

The \texorpdfstring{ZrAs$_2$}{ZrAs2} family of transition-metal dipnictides crystallizes in the orthorhombic PbCl$_2$-type structure with space group \textit{Pnma} (No.~62), a nonsymmorphic space group containing glide mirror and screw symmetries (see FIG.~\ref{fig4} (a)). These nonsymmorphic symmetry operations impose additional band-degeneracy constraints beyond inversion ($\mathcal{P}$) and time-reversal ($\mathcal{T}$) symmetry and play a decisive role in stabilizing nodal-line crossings in this material class (see FIG.~\ref{fig4} (e-g)). In contrast to accidental band crossings arising purely from band inversion, the nodal lines in \texorpdfstring{ZrAs$_2$}{ZrAs2} are symmetry-enforced along specific high-symmetry directions of the bulk Brillouin zone (BZ)~\cite{Wadge21}.

High-resolution ARPES measurements along the $\bar{\Gamma}$--$\bar{X}$ and $\bar{S}$--$\bar{X}$ directions reveal linearly dispersing conduction and valence bands intersecting near the Fermi level and extending over finite momentum intervals, consistent with nodal-line segments rather than isolated Dirac points~\cite{Wadge21}. In particular, the crossing near the $\bar{S}$ point occurs along a glide-invariant high-symmetry line. Because glide symmetry squares to a momentum-dependent phase factor, Bloch states along this line carry distinct glide eigenvalues. Bands belonging to distinct glide eigenvalue representations of the double space group cannot hybridize, enforcing a symmetry-protected degeneracy even in the presence of spin--orbit coupling. This mechanism gives rise to an ``impervious'' crossing that remains gapless within experimental resolution~\cite{Wadge21}.

First-principles calculations including SOC indicate that accidental crossings away from symmetry-protected lines generally develop gaps on the order of tens of meV. For example, in HfP$_2$, SOC opens gaps of approximately $\sim 20$~meV along portions of the nodal loop~\cite{sims2020termination}. By contrast, ARPES measurements on \texorpdfstring{ZrAs$_2$}{ZrAs2} do not resolve a detectable gap at the symmetry-protected crossing within an energy resolution of $\sim 15$--$20$~meV~\cite{Wadge21}, implying that any possible SOC-induced gap is smaller than this experimental threshold. This comparison establishes a hierarchy within the dipnictide family: partially fragile nodal loops with SOC-induced gaps~\cite{sims2020termination}, compensation-dominated nodal features~\cite{bannies2021extremely}, and symmetry-enforced SOC-robust crossings~\cite{Wadge21}.

Constant-energy contour (CEC) maps further substantiate the nodal-line character. At the Fermi level, the measured Fermi surface exhibits highly anisotropic and elongated pockets rather than circular Dirac-like contours~\cite{Wadge21}. These elongated features correspond to the intersection of extended bulk nodal loops with $E_F$. Upon increasing binding energy, the contours expand and evolve continuously, forming stripelike and rectangular features consistent with calculated projections of bulk nodal loops onto the (001) cleavage surface. Similar nodal-loop Fermi-surface geometries have also been observed in ZrP$_2$~\cite{bannies2021extremely} (FIG.~\ref{fig4} (b-d)) and HfP$_2$~\cite{sims2020termination}.

A decisive criterion for identifying bulk nodal lines is their three-dimensional ($k_z$-dependent) dispersion. Systematic photon-energy--dependent measurements in \texorpdfstring{ZrAs$_2$}{ZrAs2} reveal pronounced $k_z$ dispersion of the nodal-line bands, confirming their bulk origin~\cite{Wadge21}. In contrast, surface-derived states exhibit negligible $k_z$ dispersion.

Surface electronic structure introduces further complexity. In \texorpdfstring{ZrAs$_2$}{ZrAs2}, ARPES detects surface-localized saddle points near the Fermi level that give rise to van Hove singularities (vHs) in the surface density of states~\cite{hossain2025superconductivity}. These states exhibit negligible $k_z$ dispersion, confirming their two-dimensional character. The saddle-point dispersion enhances the density of states near $E_F$ and has been correlated with the emergence of surface-confined superconductivity exhibiting Berezinskii--Kosterlitz--Thouless (BKT) characteristics~\cite{hossain2025superconductivity}.

Taken together, ARPES investigations of the \texorpdfstring{ZrAs$_2$}{ZrAs$_{2}$} family establish the defining hallmarks of nodal-line semimetals: (i) extended linear band crossings along glide-invariant high-symmetry directions~\cite{Wadge21}, (ii) anisotropic and energy-evolving Fermi contours characteristic of nodal-loop intersections~\cite{bannies2021extremely,sims2020termination}, (iii) pronounced $k_z$ dispersion confirming bulk three-dimensionality~\cite{Wadge21}, (iv) selective robustness against SOC dictated by nonsymmorphic symmetry eigenvalues~\cite{Wadge21}, and (v) coexistence of $k_z$-invariant surface states hosting van Hove singularities~\cite{hossain2025superconductivity}. Within experimental resolution limits, \texorpdfstring{ZrAs$_2$}{ZrAs2} therefore represents a paradigmatic realization of symmetry-enforced, SOC-impervious nodal-line topology directly visualized by ARPES (see summary in the Table~\ref{tab:ZrAs2Family_ARPES}).
\begin{table}[t]
\centering
\caption{Summary of symmetry protection, nodal-line location, and ARPES signatures in representative transition-metal dipnictides of the ZrAs$_2$ family.}
\label{tab:ZrAs2Family_ARPES}

\small
\setlength{\tabcolsep}{2.5pt}
\renewcommand{\arraystretch}{1.15}

\begin{tabular}{l l @{\hspace{6pt}} l l l}
\hline\hline
Material & Symmetry & k-space & ARPES & Ref. \\
 & protection & location & signatures & \\
\hline

ZrP$_2$ 
& Nonsymmorphic 
& BZ boundary 
& Nodal loops, weak gap ($\sim$7--20 meV) 
& \cite{bannies2021extremely} \\

ZrAs$_2$ 
& Nonsymmorphic 
& High-sym. dir. 
& Robust crossings, small gaps ($\sim$2--58 meV), vHs 
& \cite{Wadge21,hossain2025superconductivity} \\

HfP$_2$ 
& Nonsymmorphic 
& BZ boundary 
& Multiple crossings, partial gap ($\sim$20 meV) 
& \cite{sims2020termination} \\

HfAs$_2$ 
& Nonsymmorphic 
& BZ boundary 
& Preserved crossings, large gap ($\sim$120 meV) 
& \cite{muhammad2024electronic} \\

\hline\hline
\end{tabular}
\end{table}

\subsection{Magnetic nodal-line semimetals}
\begin{figure}[h]
\centering
\includegraphics[width=0.9\linewidth]{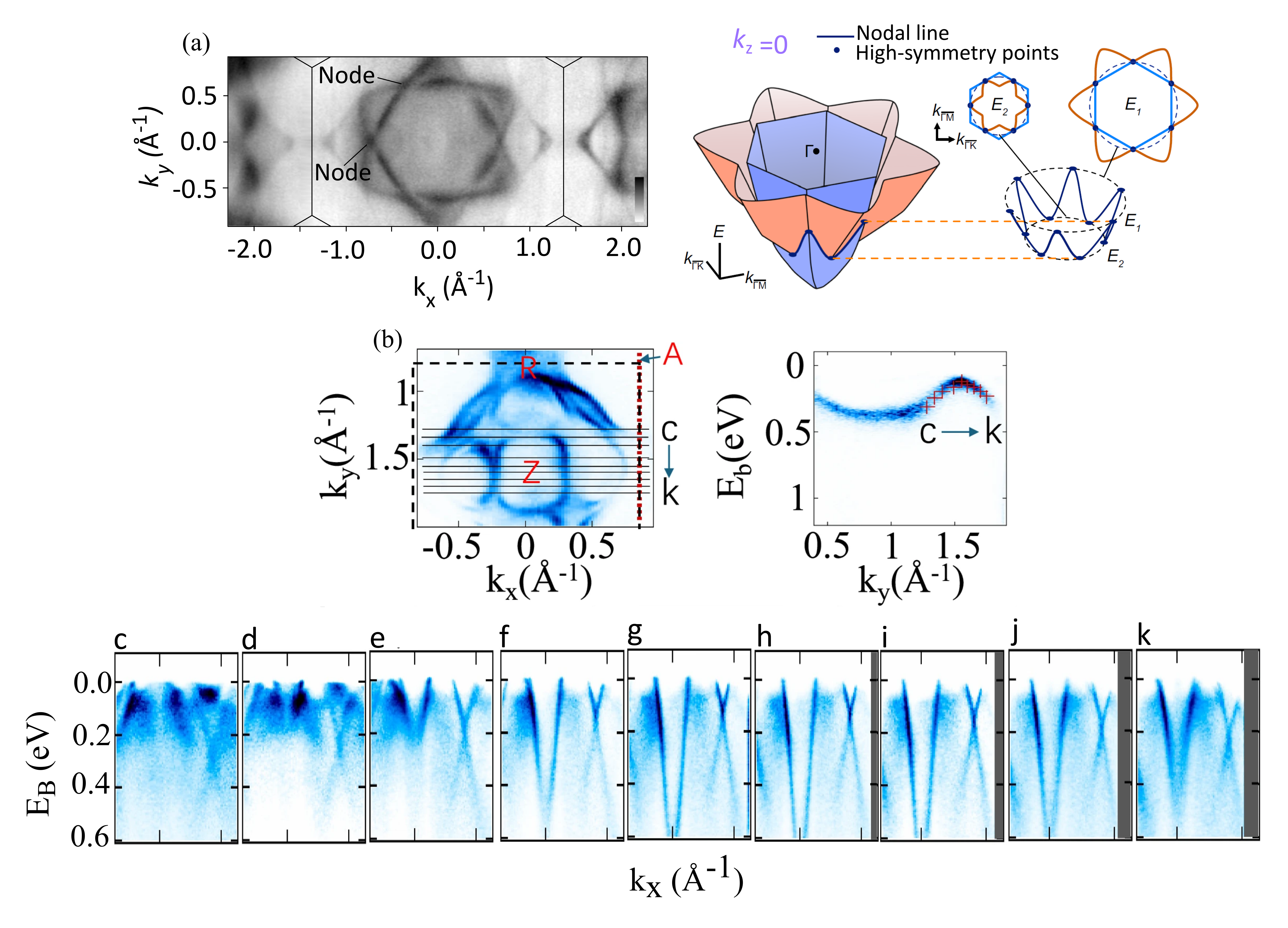}
\caption{ARPES spectra of (a) Fermi-surface map and schematic illustration of magnetic nodal lines at $k_z=0$, showing the location of symmetry-protected band crossings in elemental Co. (b) Experimental Fermi-surface intensity map with high-symmetry points
indicated, together with a representative band dispersion.
(c–k) Momentum-resolved ARPES dispersions along selected cuts, revealing the evolution of the nodal-line crossings in energy–momentum space in YMn$_2$Ge$_2$. adapted from\cite{clark2026manifold, yang2024topological}
}
\label{fig5} 
\end{figure}

ARPES has demonstrated that nodal-line band topology can persist in a wide variety of magnetic systems,
despite broken time-reversal symmetry and the presence of sizable spin--orbit
coupling and electronic correlations. From an experimental perspective,
magnetic nodal-line semimetals offer a unique opportunity to directly probe spin-polarized, symmetry-protected band crossings and their evolution in momentum space. Representative ARPES realizations include elemental
ferromagnetic Co (see FIG.~\ref{fig5} (a)) \cite{clark2026manifold}, the correlated antiferromagnet YMn$_2$Ge$_2$ (see FIG.~\ref{fig5} (b-k)) \cite{yang2024topological}, and the rare-earth $X$SbTe family \cite{hosen2018discovery,Regmi23,Regmi24,yuan2024observation,Elius25,valadez2025low,elius2025electronic},
which together span simple elemental systems, correlated intermetallics, and chemically tunable material families.
From our perspective, the nodal-line physics in elemental cobalt should not be viewed as a conventional topological feature inherited purely from lattice symmetry, but rather as a magnetically reconstructed electronic manifold that evolves across momentum space. Clark O. J. et al.~\cite{clark2026manifold} explained that the key distinction emerges when comparing different $k_z$ planes: at $k_z = 0$ ($\Gamma$-plane), continuous nodal rings enclosing high-symmetry points, forming an interconnected ``Fermi-flower''-like structure, as illustrated in FIG.~\ref{fig5}~(a)  \cite{clark2026manifold} whereas at $k_z = \pi/c$ (A-plane), the topology transforms into linear nodal lines along the A--L direction, giving rise to chain-like Dirac-cone arcs. This clearly indicates that the nodal structure is not a simple two-dimensional feature but part of a three-dimensional nodal manifold that continuously reshapes with $k_z$. Crucially, unlike time-reversal-symmetric nodal-line systems, these degeneracies persist in cobalt despite broken time-reversal symmetry, which points to the dominant role of magnetic space-group symmetries, particularly mirror eigenvalue protection in specific planes and symmetry-enforced degeneracies along high-symmetry directions. This protection is highly sensitive to the magnetization axis: symmetry lowering can lift the nodal crossings and reconstruct the bands, implying that the topology is intrinsically magnetization-dependent and tunable. Another important aspect that distinguishes cobalt is the spin-selective nature of the nodal manifold, where the crossings are predominantly carried by a single (minority) spin channel due to exchange splitting, leading to a situation where both intersecting bands share the same spin polarization. This behavior is fundamentally different from both spin-degenerate nodal lines and conventional Weyl systems, and it naturally enables reversible spin control of the nodal features under external magnetic fields. Experimentally, these characteristics manifest as distinct ARPES signatures, including closed intensity contours at $k_z = 0$ (see FIG.~\ref{fig5} (a)) and dispersive arc-like features at $k_z = \pi/c$, consistent with the evolution from ring-like to line-like topology. Overall, we interpret these observations as evidence that elemental cobalt hosts a robust, spin-polarized nodal-line manifold governed by magnetic symmetry, where topology, spin, and momentum are strongly intertwined, establishing a platform that goes beyond conventional nodal-line semimetals and opens a route toward controllable topological states in simple ferromagnetic systems~\cite{clark2026manifold}.

The emergence of the Dirac nodal line in YMn$_2$Ge$_2$ is governed by a nontrivial interplay of magnetic and nonsymmorphic symmetries. Although conventional time-reversal symmetry $\mathcal{T}$ is broken due to antiferromagnetic ordering, the system preserves a combined antiunitary space--time inversion symmetry $P\tilde{T}$, where $\tilde{T}$ denotes time reversal followed by a half-lattice translation. This symmetry enforces Kramers-like double degeneracy throughout the Brillouin zone. In addition, the presence of nonsymmorphic glide mirror symmetries further constrains the band structure by assigning distinct symmetry eigenvalues to the bands. Along the Brillouin-zone boundary lines, the algebraic relations between these glide operations and $P\tilde{T}$ enforce unavoidable crossings between doubly degenerate bands, resulting in robust fourfold-degenerate Dirac nodal lines\cite{yang2024topological}. 
As confirmed by momentum-resolved ARPES measurements and theoretical analysis, these nodal crossings extend continuously along high-symmetry directions, forming a symmetry-protected nodal-line manifold rather than isolated Dirac points. The Fermi surface in Fig.~\ref{fig5}~(b) shows clear spectral features along the C--K direction, where the band dispersion exhibits Dirac-like crossings. These nodes appear at different momenta between C and K and shift continuously in energy and momentum, indicating they are part of an extended nodal line rather than isolated points. This is further confirmed by the cuts in Fig.~\ref{fig5}~(c--k), which traces the evolution of the crossings and demonstrates their connectivity along the high-symmetry direction. These features are therefore not accidental but symmetry-enforced, and they remain stable as long as both $P\tilde{T}$ and the relevant glide mirror symmetries are preserved, while breaking either of them leads to lifting of the degeneracy and gap opening. 
The research on YMn$_2$Ge$_2$ \cite{yang2024topological} opens new pathways for ARPES-based investigations within the broader RMn$_2$Ge$_2$ family, where variations in magnetic order and electronic correlations may lead to diverse nodal and Weyl states. For example, compounds such as SmMn$_2$Ge$_2$ and TbMn$_2$Ge$_2$, which host distinct magnetic ground states, are promising candidates to explore the evolution of nodal topology and possible Weyl features. However, systematic ARPES measurements on these materials remain largely unexplored, and such studies are expected to provide direct momentum-resolved insight into the interplay of magnetism, topology, and Hund-driven correlations across the family.

The rare-earth $X$SbTe ($X$ = Pr, Nd, Gd, Tb, Dy, Er) family represents a chemically tunable platform in which ARPES can systematically track the evolution of nodal-line–derived electronic structure across a magnetic series (see FIG.~\ref{fig2} (e)) \cite{hosen2018discovery,Regmi23,Regmi24,
yuan2024observation,Elius25,valadez2025low,elius2025electronic}. Across multiple members of this family, ARPES measurements reveal nodal-line–derived band crossings along high-symmetry directions of the bulk Brillouin zone, most prominently along the $\bar{X}-\bar{R}$ and $\bar{\Gamma}-\bar{M}$ directions. A recurring experimental hallmark is the appearance of a diamond-shaped Fermi surface centered at the $\bar{\Gamma}$ point, originating from Dirac-like crossings that form nodal lines or nodal planes in momentum space.

In lighter rare-earth compounds such as PrSbTe and NdSbTe, ARPES spectra resolve predominantly gapless nodal-line features that closely resemble those observed in nonmagnetic ZrSiS-type systems \cite{Regmi24,Regmi23,yuan2024observation},
indicating that the presence of magnetic moments alone does not immediately destroy the nodal-line band topology. In GdSbTe and TbSbTe, similar nodal-line–derived crossings persist \cite{hosen2018discovery,Elius25}, while subtle momentum-dependent modifications and partial gapping become visible at higher binding energies, reflecting an increasing influence of spin--orbit coupling. In the heaviest members of the series, DySbTe and ErSbTe, ARPES measurements provide direct evidence for a pronounced spin--orbit–induced gap
opening along directions that host nodal lines in lighter compounds, signalling a transition toward gapped topological phases from an experimental standpoint \cite{valadez2025low,elius2025electronic}.

In addition to bulk nodal-line features, ARPES measurements on $X$SbTe
compounds frequently reveal surface and near-surface electronic states
\cite{Regmi24,yuan2024observation}. Weakly dispersive
surface-localized bands confined to the surface projection of the bulk nodal
lines are observed in several members of the series. While not all studies
explicitly identify these states as topological drumhead surface states, their
limited dispersion and surface sensitivity are consistent with
nodal-line–derived surface features, highlighting the importance of carefully
distinguishing bulk and surface contributions in magnetic nodal-line systems.

Overall, ARPES studies on elemental Co, YMn$_2$Ge$_2$, and the $X$SbTe family demonstrate that nodal-line band topology can remain robust even in the absence of time-reversal symmetry, provided it is protected by appropriate crystalline or magnetic symmetries. In particular, nodal lines are stabilized either by mirror or nonsymmorphic glide symmetries, or by composite antiunitary symmetries such as $P\tilde{T}$, which enforce Kramers-like degeneracy in antiferromagnets. While glide symmetries typically confine nodal lines to high-symmetry directions or Brillouin-zone boundaries, their experimental manifestation in ARPES varies depending on spin--orbit coupling, electronic correlations, and magnetic configuration. Notably, ferromagnetic systems exhibit a strong dependence on the magnetization direction, whereas antiferromagnets with $P\tilde{T}$ symmetry host more robust nodal features. The symmetry conditions and corresponding design principles, along with their experimentally observed ARPES signatures, are summarized in Tables~\ref{MagneticNLSM_symmetry} and~\ref{magarpes}.
\begin{table}[t]
\centering
\caption{Symmetry conditions and design principles for magnetic nodal-line semimetals.}
\label{MagneticNLSM_symmetry}

\small
\setlength{\tabcolsep}{4pt}
\renewcommand{\arraystretch}{1.1}

\begin{tabular}{l l l l l}
\hline\hline
Material & Order ($T_{c}/T_{N}$) & Symmetry protection & k-space location & Mag. dep. \\
\hline

Co 
& FM ($\sim1388~K$)
& Mirror ($M_z$) 
& Mirror planes 
& Strong \\

YMn$_2$Ge$_2$ 
& AFM ($\sim395~K$)
& $P\tilde{T}$ + glide 
& BZ boundary (A--R) 
& Weak \\

PrSbTe 
& AFM ($\sim10-15~K$)
& Glide 
& BZ boundary 
& Moderate \\

NdSbTe 
& AFM ($\sim2-5~K$)
& $P\tilde{T}$ + glide 
& BZ boundary 
& Weak--mod. \\

GdSbTe 
& AFM ($\sim12~K$)
& $P\tilde{T}$ 
& BZ boundary 
& Weak \\

TbSbTe 
& AFM ($\sim7~K$)
& Glide 
& BZ boundary 
& Moderate \\

DySbTe 
& AFM ($\sim3-5~K$)
& Glide 
& BZ boundary 
& Moderate \\

ErSbTe 
& AFM ($\sim2-3~K$)
& Glide 
& BZ boundary 
& Moderate \\

\hline\hline
\end{tabular}
\end{table}
\begin{table}[t]
\centering
\caption{Key ARPES signatures of magnetic nodal-line semimetals.}
\label{magarpes}

\small
\setlength{\tabcolsep}{4pt}
\renewcommand{\arraystretch}{1.1}

\begin{tabular}{l l l}
\hline\hline
Material & Key ARPES signatures & Ref. \\
\hline

Co 
& Spin-polarized nodal crossings, Fermi-surface loops 
& \cite{clark2026manifold} \\

YMn$_2$Ge$_2$ 
& Fourfold Dirac crossings, nodal dispersion, drumhead states 
& \cite{yang2024topological} \\

PrSbTe 
& Dirac-like crossings, weak nodal dispersion 
& \cite{Regmi24,yuan2024observation} \\

NdSbTe 
& Multiple nodal lines, symmetry-enforced crossings 
& \cite{Regmi23} \\

GdSbTe 
& Robust crossings across $T_N$, surface states 
& \cite{hosen2018discovery} \\

TbSbTe 
& Linear crossings, SOC-induced modifications 
& \cite{Elius25} \\

DySbTe 
& Gap opening due to SOC 
& \cite{valadez2025low} \\

ErSbTe 
& Transition toward gapped nodal features 
& \cite{elius2025electronic} \\

\hline\hline
\end{tabular}
\end{table}
\newpage

    
\section{Macroscopic signature in transport coefficients}
As seen in previous sections, in nodal-line semimetals, the conduction and valence bands touch each other along lines or loops in momentum space. These band crossings are not guaranteed by translational symmetry alone but require additional crystalline symmetries, such as mirror or inversion symmetry. When the nodal line lies at or near the Fermi level, it can strongly influence transport properties, as the low-energy excitations are dominated by the linearly dispersing states around the nodal line. This topologically protected band crossing gives rise to distinctive electronic phenomena, including drumhead-like surface states, a geometric phase of the electronic wavefunction manifested in quantum oscillations, non-saturating magnetoresistance, and quantum interference effects such as weak localization or antilocalization. In cases where time-reversal symmetry is broken, the nodal line can enhance the intrinsic anomalous Hall effect (AHE). Moreover, the nodal line can enable macroscopic realizations of quantum anomalies, including the chiral anomaly. The resulting transport signatures and their experimental manifestations are discussed in the following section.

\subsection{ Quantum oscillations}
The application of a magnetic field quantizes energy bands, creating what are known as Landau tubes in k-space. As the magnetic field increases, these Landau tubes expand, resulting in a peak in the density of electronic states when they cross the Fermi level. This situation is manifested as quantum oscillations with the inverse magnetic field. The quantum oscillations in magnetization are known as the de Haas–van Alphen (dHvA) effect, while the quantum oscillations in resistivity are referred to as the Shubnikov–de Haas (SdH) effect. In this review, we focus solely on the SdH effect.

The fundamental oscillation frequency $F$ (in tesla) is related to the extremal cross-sectional area of the Fermi surface $A_{\mathrm{ext}} \equiv A(E_F)$ normal to the magnetic field $B$ via the Onsager relation:
\begin{equation}
F \;=\; \frac{\hbar}{2\pi e}\,A_{\mathrm{ext}},
\qquad\Longleftrightarrow\qquad
A_{\mathrm{ext}} \;=\; \frac{2\pi e}{\hbar}\,F.
\label{eq:onsager}
\end{equation}
Only extremal orbits contribute because integrating over the momentum component along the magnetic field direction leads to destructive interference, except at the stationary points of \( A(k_\parallel) \) (where the cross sections are maximal or minimal). Extreme orbits, which are closed orbits on the Fermi surface and are perpendicular to the applied magnetic field, contribute to quantum oscillations. As a result, these oscillations can reveal the shape and size of the Fermi surface, as well as whether a given orbit is closed or open. Quantum oscillations are now widely used to reconstruct the Fermi surface of materials. Additionally, the phase accumulated by electrons during cyclotron motion on closed orbits results in a phase shift in the quantum oscillation pattern. Therefore, analyzing this phase shift provides direct information about the topology of the electronic band structure \cite{mikitik1999manifestation}.

\begin{figure}[h]
\centering
\includegraphics[width=0.9\linewidth]{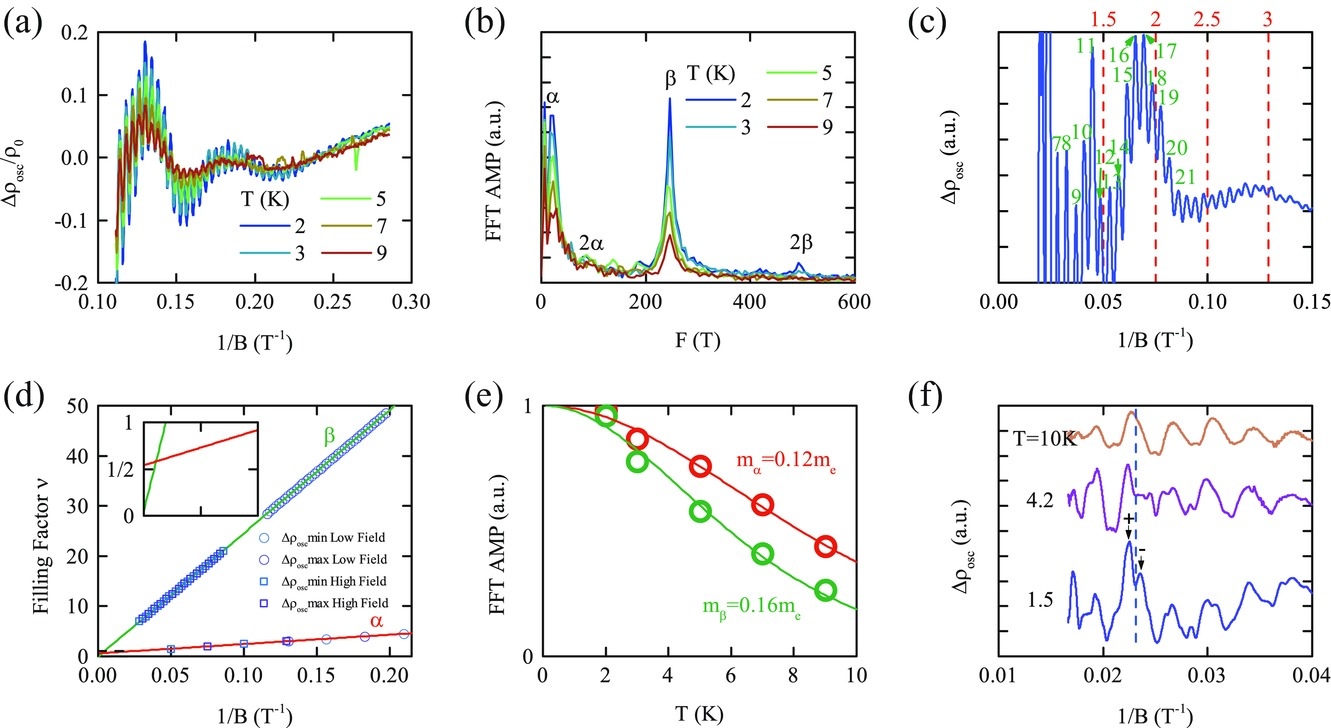}
\caption{ SdH oscillations and associated nontrivial Berry phase in ZrSiS:  
(a) Relative MR oscillations versus inverse magnetic field (1/$B$).  
(b) Corresponding FFT shows two peaks reflecting two Fermi pockets.
(c) Oscillations in a pulsed magnetic field of  6 to 50 T at 2 K  show $\alpha$ and $\beta$ modes, with Landau level indices marked in red and green.  
(d) Landau fan diagrams for the two SdH modes and intercepts from linear extrapolation, the inset shows an enlargement near zero. The low-frequency $\alpha$ mode indicates a nontrivial Berry phase.  
(e) Amplitude analysis for effective mass extraction.  
(f) Oscillating MR at various temperatures, indicating energy splitting. Figure adapted from reference \cite{wang2016evidence}.
}
\label{fig6}
\end{figure}

\begin{figure}[h]
\centering
\includegraphics[width=0.9\linewidth]{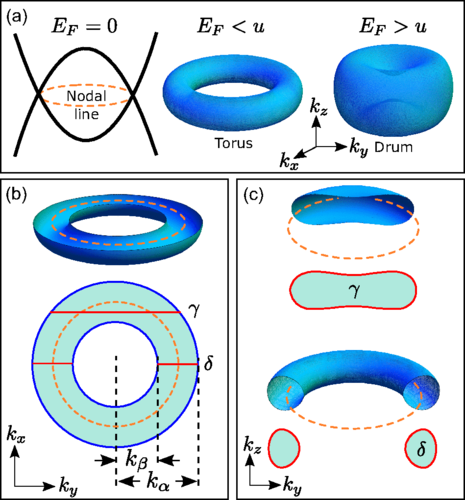}
\caption{(a) The nodal line, torus, and drum Fermi surfaces for
a nodal-line semimetal, wherein the dashed ring indicates the nodal ring. $E_F$ is the Fermi energy, and $u$ is a model parameter. (b) The maximum ($\alpha$) and minimum ($\beta$) cross sections in the nodal-line plane of the torus Fermi surface. (c) The maximum ($\gamma$) and minimum ($\delta$) cross sections out of the nodal-line plane of the torus Fermi surface. The figure is adapted from reference~ \cite{li2018rules}.
}
\label{fig7}
\end{figure}

In a nodal-line semimetal, an electron acquires a nontrivial $\pi$ Berry phase when moving around a loop that encloses the nodal line. In contrast, it has a trivial Berry phase around a loop parallel to the nodal ring plane \cite{Fang15}. The characteristic nontrivial Berry phase, as mentioned earlier, is probed by quantum oscillations. 
In the SdH effect, the oscillatory signal originates from the oscillatory density of states at the Fermi level and
appears most naturally in resistivity, given by the Lifshitz-Kosevich formula: 
$\rho \sim \cos\!\left[2\pi\left(\frac{F}{B}+\phi\right)\right]$, where B is the magnetic field, F is the oscillation frequency and $\phi$ is the phase shift, which can provide the crucial information about the band topology. Experimentally, $\phi$ is extracted from a fan diagram by fitting $n^{th}$ peaks in resistivity in a magnetic field with a linear function: n=F/B+$\phi$. Here, F is the slope of the linear function and $\phi$ is the intercept on the n-axis. 
Recently, non-magnetic nodal-line semimetals have been observed to exhibit quantum oscillations with a non-trivial Berry phase due to the breaking of the inversion symmetry. Indeed, a finite Berry curvature and the associated Berry phase emerge when either time-reversal symmetry or spatial inversion symmetry is broken, whereas the simultaneous presence of both symmetries forces the Berry curvature to vanish throughout momentum space. A particular study highlights these quantum oscillations along with a non-trivial Berry phase in ZrSiS, shown in FIG.~\ref{fig6} \cite{wang2016evidence}.
To understand the origin of the non-trivial Berry phase, we should first discuss the shape of the Fermi surface in nodal-line semimetals and examine whether the extreme cross-section encloses the nodal line or point. Most nodal-line semimetals have a torus-shaped Fermi surface, as shown in FIG.~\ref{fig7}  ($E_F< \mu$, where $E_F$ is the Fermi energy and $u$ is a model parameter)\cite{li2018rules}. 

The phase shift analysis in nodal-line semimetals is more complex due to the toroidal Fermi surface, which allows for both maximum and minimum cross-sections, unlike the spherical Fermi surface that only has a maximum extreme cross-section. As mentioned earlier, the quantum oscillations probed in a magnetic field that is perpendicular to the extreme area of cross-section enclosing the nodal line exhibit a non-trivial Berry phase; otherwise, they would have a trivial Berry phase. Here, along the circle of the $\alpha$ and $\beta$, the Berry phase is 0, leading to a phase shift value 5/8 (for holes) or -5/8 (for electrons), whereas, along the circle $\delta$, the Berry phase is $\pi$, resulting in a phase shift of 1/8 (for electrons) and -1/8 (for holes). The results are summarized in Table \ref{tab:quantum oscillations}, which is reproduced from the literature~\cite{li2018rules}.
\begin{table}[h!]
\centering
\begin{tabular}{c c c c c}
\hline\hline
\textbf{Cross section} & \textbf{Berry phase} & \textbf{Min./Max} &
\textbf{Electron} & \textbf{Hole} \\
\hline
$\alpha$ & $0$ & Max. & $-1/2 + 0 - 1/8 = -5/8$ & $+5/8$ \\
$\beta$  & $0$ & Min. & $-1/2 + 0 + 1/8 = -3/8 \leftrightarrow 5/8$ & $-5/8$ \\
$\gamma$ & $0$ & Max. & $-1/2 + 0 - 1/8 = -5/8$ & $+5/8$ \\
$\delta$ & $\pi$ & Min. & $-1/2 + \pi/2\pi + 1/8 = 1/8$ & $-1/8$ \\
\hline\hline
\end{tabular}
\caption{Phase shifts of quantum oscillations associated with the extremal 
cyclotron orbits ($\alpha$, $\beta$, $\gamma$, and $\delta$) of the torus-shaped Fermi surface in a nodal-line semimetal for electrons and holes. The symbols $\alpha$ and $\beta$ correspond to the outer (maximum) and inner (minimum) cross sections in the nodal-line plane, while $\gamma$ and $\delta$ denote extremal cross sections away from the nodal-line plane. ``Min.'' and ``Max.'' indicate minimum and maximum extremal cross sections, respectively.}
\label{tab:quantum oscillations}
\end{table}


\subsection{The weak (anti-)localization effect}
\begin{figure}[h]
\centering
\includegraphics[width=0.6\linewidth]{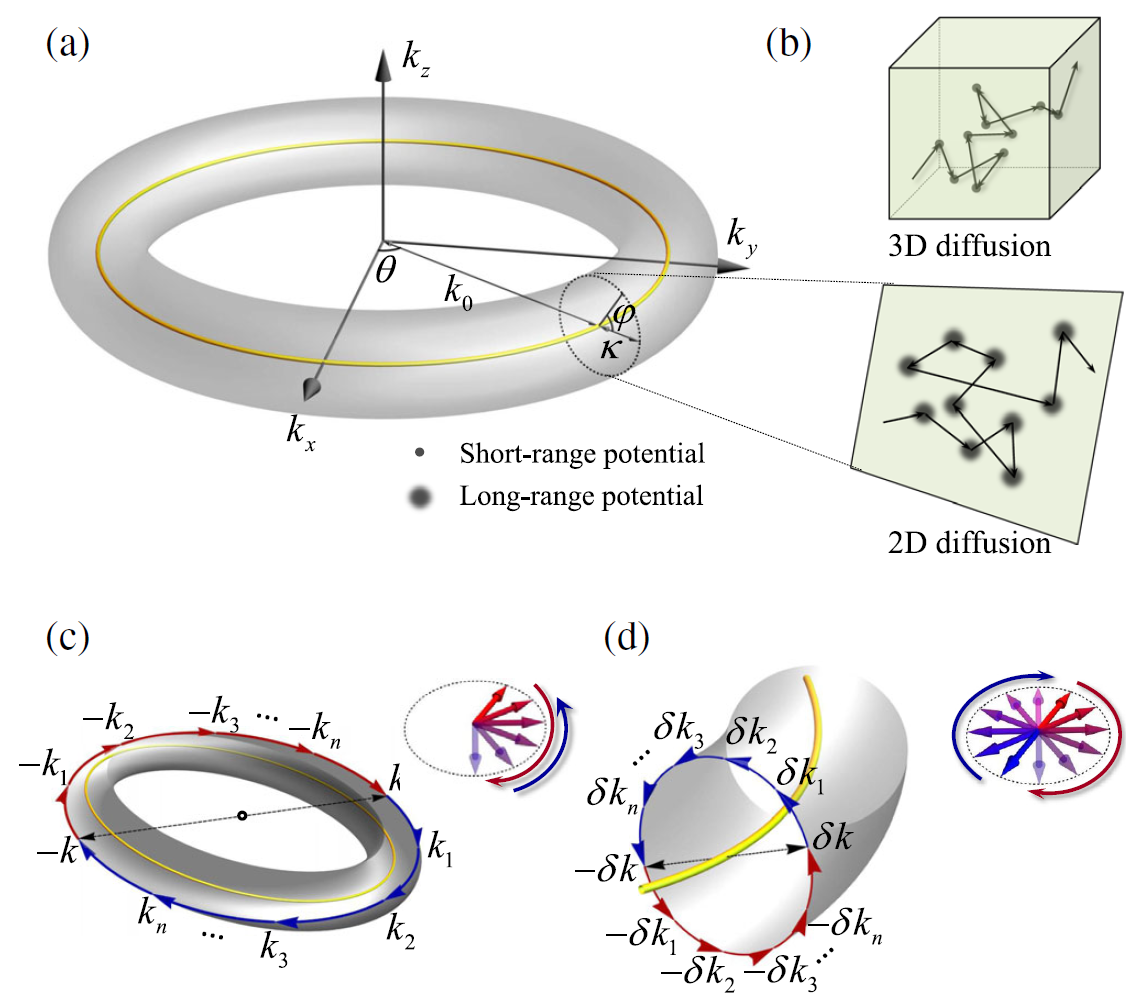}
\caption{Backscattering processes.  
(a) Torus-shaped Fermi surface with major radius \( k_0 \), minor radius \( \kappa \), and angles \( \theta \) (toroidal) and \( \phi \) (poloidal).  
(b) Short-range impurities lead to 3D diffusion; long-range ones cause 2D diffusion.  
(c) Short-range impurities result in coherent backscattering along the toroidal direction with zero net spinor rotation.  
(d) Long-range impurities cause backscattering in the poloidal direction, leading to a \( 2\pi \) spinor rotation and \( \pi \) Berry phase, resulting in weak antilocalization. This figure was adapted from reference~ \cite{chen2019weak}.
}
\label{fig8}
\end{figure}

Weak localization (WL) effect is a quantum-mechanical phenomenon in which the interplay of disorder creates a unique situation. In this realm, constructive interference among various scattering events results in a remarkable increase in backscattering \cite{lee1985disordered}. Such an effect is anticipated at very low temperatures; the electronic system is expected to exhibit weak disorder in the quantum diffusive regime, where the mean free path is significantly smaller than the phase-coherence length. In such a regime, disorder-induced quantum interference effects become significant \cite{lee1985disordered}. On the other hand, the suppression of backscattering is associated with the weak antilocalization (WAL) effect \cite{kawaguchi1978angular, hikami1980spin}. 
Both Weak localization and weak antilocalization lead to additional quantum corrections to the electrical conductivity. Moreover, the non-trivial topology of band structures enriches the conventional result of quantum diffusion, establishing the fascinating interplay between the diffusion dimensionality and the band topology. Specifically, the interference loop encircles the nodal line, which carries a Berry phase of $\pi$, and this phase plays a crucial role in governing the quantum diffusion \cite{chen2019weak}. In a nodal-line semimetal, the torus-shaped Fermi surface creates additional backscattering channels relative to the conventional spherical Fermi surface, thereby strongly modifying quantum diffusion behavior. There are two backscattering channels for the torus-shaped Fermi surface. The type of scattering dominates the WL or WAL effects depending on the scattering range of disorder potential, long-range (LR) and short-range (SR) disorder, relative to the nodal loop \cite{chen2019weak}, as shown in FIG.~\ref{fig8}.

\begin{figure}[h]
\centering
\includegraphics[width=0.9\linewidth]{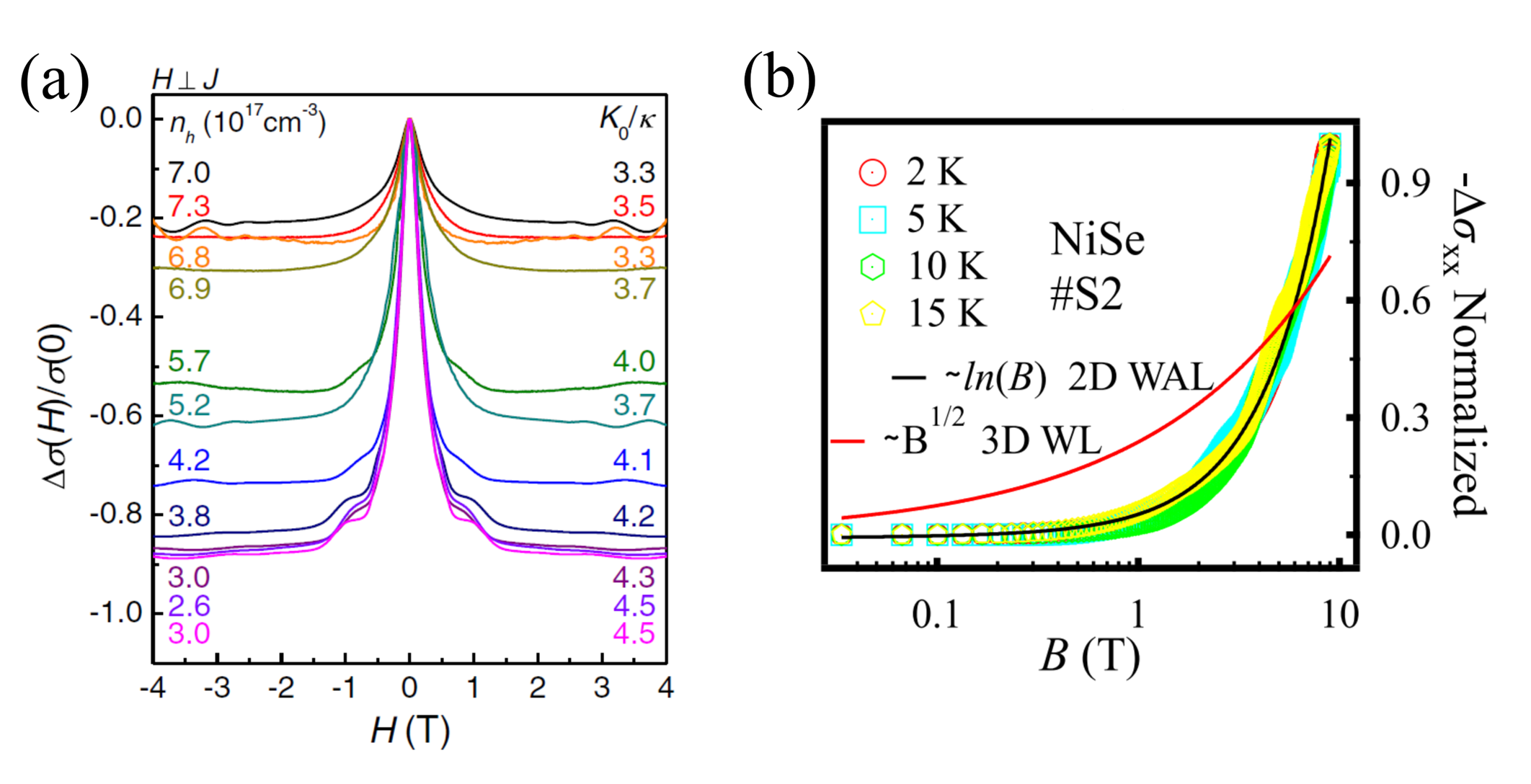}
\caption{(a) Weak antilocalization of nodal-line fermions in SrAs$_3$: The low-field transverse magnetoconductivity, represented as $\Delta\sigma(H)/\sigma(0)$ at a temperature of 2 K and under the condition where the magnetic field $H$ is perpendicular to the current density $J$, is analyzed for eleven SrAs$_3$ crystals.These crystals exhibit varying hole densities $n_h$ and ratios of $K_0/\kappa$, where $K_0$ and $\kappa$ are the major and minor radii of the torus-shaped Fermi surface, respectively. This figure is adapted from reference \cite{kim2022quantum}, and (b) shows normalized magnetoconductivity. The black line shows 2D WAL fitting, and the red line is for 3D WL fitting. This figure is adapted from reference \cite{pradhan2024topological}.
}
\label{fig9} 
\end{figure}

An applied magnetic field suppresses quantum corrections and induces dephasing, thereby enabling probing of the WL and WAL effects in magnetoconductivity measurements. For example, in the SR regime, diffusion is 3D, and the WL effect leads to positive magnetoconductivity, whereas in the LR regime, WAL leads to negative magnetoconductivity \cite{yang2022quantum}. 
In most of the nodal-line semimetals, the disorder potential is expected to be long-range due to unconventional screening effects \cite{syzranov2017electron}. For instance, in slightly hole-doped SrAs$_3$, two-dimensional behavior of WAL has been observed, namely a sharp peak in $\Delta \sigma(H)$ \cite{kim2022quantum}, which is reported in FIG.~\ref{fig9}(a). Furthermore, in the NiSe single crystal, an increasing negative value of $\sigma_{xx}$ with increasing B has been characterized as a consequence of the weak antilocalization effect. Additionally, weak-field magnetoconductivity was well fitted with $-\ln(B)$, as predicted for scaling behavior in two-dimensional WAL \cite{pradhan2024topological}. See the figure below, FIG.~\ref{fig9}(b). In Mg$_3$Bi$_2$, weak antilocalization has been explained by the nodal-line semimetal model \cite{zhou2020quantum}. Therefore, these transport signatures of WAL serve as a characteristic of the nodal line due to the $\pi$ Berry phase.

\subsection{Extremely large non-saturating magneto-resistance}

\begin{figure}[h]
\centering
\includegraphics[width=0.6\linewidth]{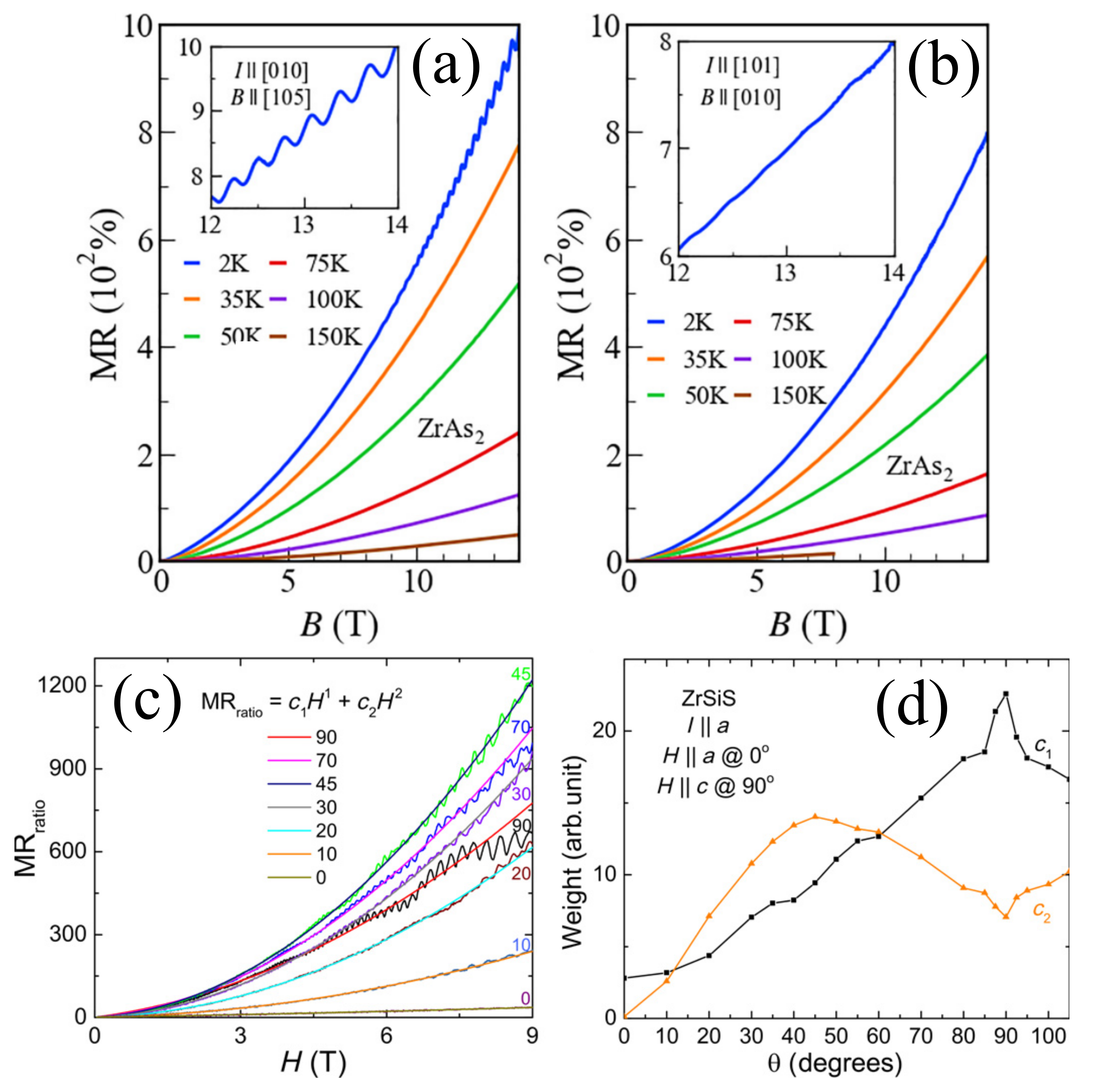}
\caption{(a), (b) Non-saturating magnetoresistance (MR) in nodal-line semimetal ZrAs$_2$ at various temperatures for the current \(\mathbf{I \parallel [010]}\) and \(\mathbf{I \parallel [101]}\), respectively. The insets show the corresponding quantum oscillations in the 12--14~T magnetic-field range. This figure is adapted from reference\cite{nandi2024magnetotransport}, whereas MR at various angles in ZrSiS. 
(c) MR vs. \(\mu_{0}H\) at various angles with fits to \(\mathrm{MR} = 1 + c_{1}H + c_{2}H^{2}\). Solid lines are fits; oscillatory curves are data. (d) Extracted \(c_{1}\) and \(c_{2}\) from 9-T MR curves at angles from \(0^\circ\) to \(105^\circ\); squares and triangles represent \(c_{1}\) and \(c_{2}\), respectively. This figure is adapted from reference \cite{Ali2016SciAdv}.
}
\label{fig10} 
\end{figure}

The variation in electrical resistance in the presence of an external magnetic field is referred to as magnetoresistance. Conventional metals and semimetals show a quadratically increasing magnetoresistance that eventually saturates at sufficiently large magnetic fields for $(\mu B \gg 1)$, where $\mu$ is the mobility and $B$ is the magnetic field.
In contrast, the topological semimetals exhibit bands with linear dispersion, which leads to modulated electronic accumulation in the lowest Landau level, resulting in linear or quadratic non-saturating magnetoresistance in high magnetic fields \cite{abrikosov1998quantum}. Such non-saturating linear MR is viewed as characteristic of the macroscopic transport signature of non-trivial topology originated from large Berry curvature associated with linear bands, reflecting into an extremely large MR ratio\cite{Shekhar2015,Campbell2021}. However, the quadratic non-saturating MR could also result from the complete compensation of charge carriers \cite{ali2014large}. Otherwise, in the case of an open Fermi surface or orbit, the MR does not saturate. Therefore, MR measurements give useful information whether the Fermi surface is closed or open\cite{peierls1955quantum}. Nodal-line semimetals are observed to exhibit the large non-saturating MR as shown in FIG.~\ref{fig10} (a, b)\cite{nandi2024magnetotransport}. The trend is either quadratic or linear, depending on the origin, complete compensation of charge carriers, or non-trivial topology. For example, Singha et al. observed linear magnetoresistance in high magnetic field regions \cite{singha2017large}. A recent theoretical study has captured the experimentally measured magnetoresistance in ZrSiX (X= S, Se, Te) nodal-line semimetals, FIG.~\ref{fig10} (c, d) \cite{Ali2016SciAdv, wang2016evidence} through their calculations, including the unusual butterfly-shaped anisotropic magnetoresistance \cite{zhang2025magnetoresistance}.

\subsection{Anomalous Hall effect}

The anomalous Hall effect (AHE) is a well-known phenomenon in materials with broken time-reversal symmetry (such as ferromagnets and altermagnets), where a transverse voltage is generated by a longitudinal current. This effect originates from the interplay of breaking of time-reversal symmetry and strong spin–orbit coupling\cite{Nagaosa2010RMP}. In recent years, the intrinsic contribution to the AHE has been linked to a non-zero Berry curvature: a topological property that behaves like a fictitious magnetic field in momentum space and arises from non-trivial band structures \cite{Xiao2018PRX,sakai2018giant, manna2018heusler,PhysRevLett.127.127202}. However, the AHE can also originate from extrinsic mechanisms such as skew scattering and side jump.

In the skew‑scattering mechanism, asymmetric impurity scattering caused by effective spin–orbit coupling of either electrons or impurities produces an additional transverse contribution to the Hall voltage. On the other hand, in the side‑jump mechanism, the electron experiences opposite effective electric fields as it approaches and then leaves an impurity, causing its velocity to deflect in opposite directions during the two stages. The time‑integrated velocity deflection in this process constitutes the side‑jump displacement.

Based on previous experimental observations and theoretical predictions, the behavior of the AHE as a function of longitudinal conductivity $\sigma_{xx}$ can be categorized into at least three regimes:
(i) \emph{High‑conductivity regime} [$\sigma_{xx} > 10^{6}\,(\Omega\!\cdot\!\mathrm{cm})^{-1}$], where skew scattering dominates and $\sigma_{xy} \propto \sigma_{xx}$;
(ii) \emph{Intermediate regime} [$10^{4} < \sigma_{xx} < 10^{6}\,(\Omega\!\cdot\!\mathrm{cm})^{-1}$], where the intrinsic AHE remains nearly constant;
(iii) \emph{Low‑conductivity regime} [$\sigma_{xx} < 10^{4}\,(\Omega\!\cdot\!\mathrm{cm})^{-1}$], where $\sigma_{xy}$ decreases faster than linearly with a scaling exponent of 1.6–1.7.
The origin of this behavior remains an open problem \cite{Nagaosa2010RMP}.

The relationship between anomalous Hall conductivity ($\sigma_{xy}$) and longitudinal conductivity ($\sigma_{xx}$) helps identify the underlying physical mechanisms. For example, when skew scattering dominates, $\sigma_{xy} \propto \sigma_{xx}$. In contrast, for side‑jump contributions, $\sigma_{xy} \propto \sigma_{xx}^{2}$, which depends on residual resistivity or conductivity and is largely temperature‑independent.

With this foundation, we now discuss the AHE in \emph{nodal‑line semimetals} and the relevant mechanisms observed in these systems. In magnetic nodal lines, where energy bands touch along a line or loop, a large Berry curvature can arise \cite{Ali2016SciAdv}, resulting in an anomalous Hall conductivity (AHC) \cite{gangwar2025evidences,Chatterjee_2023}. In the magnetic nodal‑line semimetal MnAlGe, the AHC is significantly enhanced to about $700~(\Omega\!\cdot\!\mathrm{cm})^{-1}$ \cite{guin20212d}. The intrinsic AHE is confirmed through scaling of $\rho^{A}_{yx}$ versus $\rho_{xx}^{2}$, and the zero‑field conductivity of $\sim 10^{4}~(\Omega\!\cdot\!\mathrm{cm})^{-1}$ places MnAlGe within the intrinsic regime. Here, nodal‑line band anticrossings near the Fermi level generate large Berry curvature, producing one of the highest reported AHCs.

In the nodal‑surface semimetal Fe$_3$Ge, a giant AHC of $1500~(\Omega\!\cdot\!\mathrm{cm})^{-1}$ has been reported and attributed to large Berry curvature generated by a gapless nodal surface \cite{li2025giant}. In Fe$_3$GeTe$_2$, the scaling behavior of $\sigma_{xy}$ follows $\sigma_{xy} \propto \sigma_{xx}^{1.7}$, and a large $\sigma_{xy}$ of $\sim 180~(\Omega^{-1}\,\mathrm{cm}^{-1})$, both of which are attributed to intrinsic Berry curvature originating from nodal lines \cite{Kim2018NatMater}. In Mn$_3$GaC, the AHE arises from a combination of intrinsic Berry‑phase contributions and extrinsic skew‑scattering mechanisms. Additionally, the Berry curvature significantly enhances the Nernst effect, generating a large Nernst coefficient under an applied temperature gradient \cite{gangwar2025evidences}.

Overall, the interplay between Berry curvature, nodal‑line topology, and SOC provides promising opportunities for optimizing electrical and thermoelectric responses in magnetic materials, with implications for spintronics and advanced sensing technologies.

\subsection{Chiral anomaly} 
 The chiral anomaly is the breakdown of the conservation law of chirality at the quantum level. The chiral anomaly, also known as the Adler–Bell–Jackiw anomaly, was initially proposed in relativistic quantum field theory \cite{nielsen1983adler}. It describes the creation and annihilation of relativistic particles with well-defined chirality or handedness. 
 
\begin{figure}[h]
\centering
\includegraphics[width=0.7\linewidth]{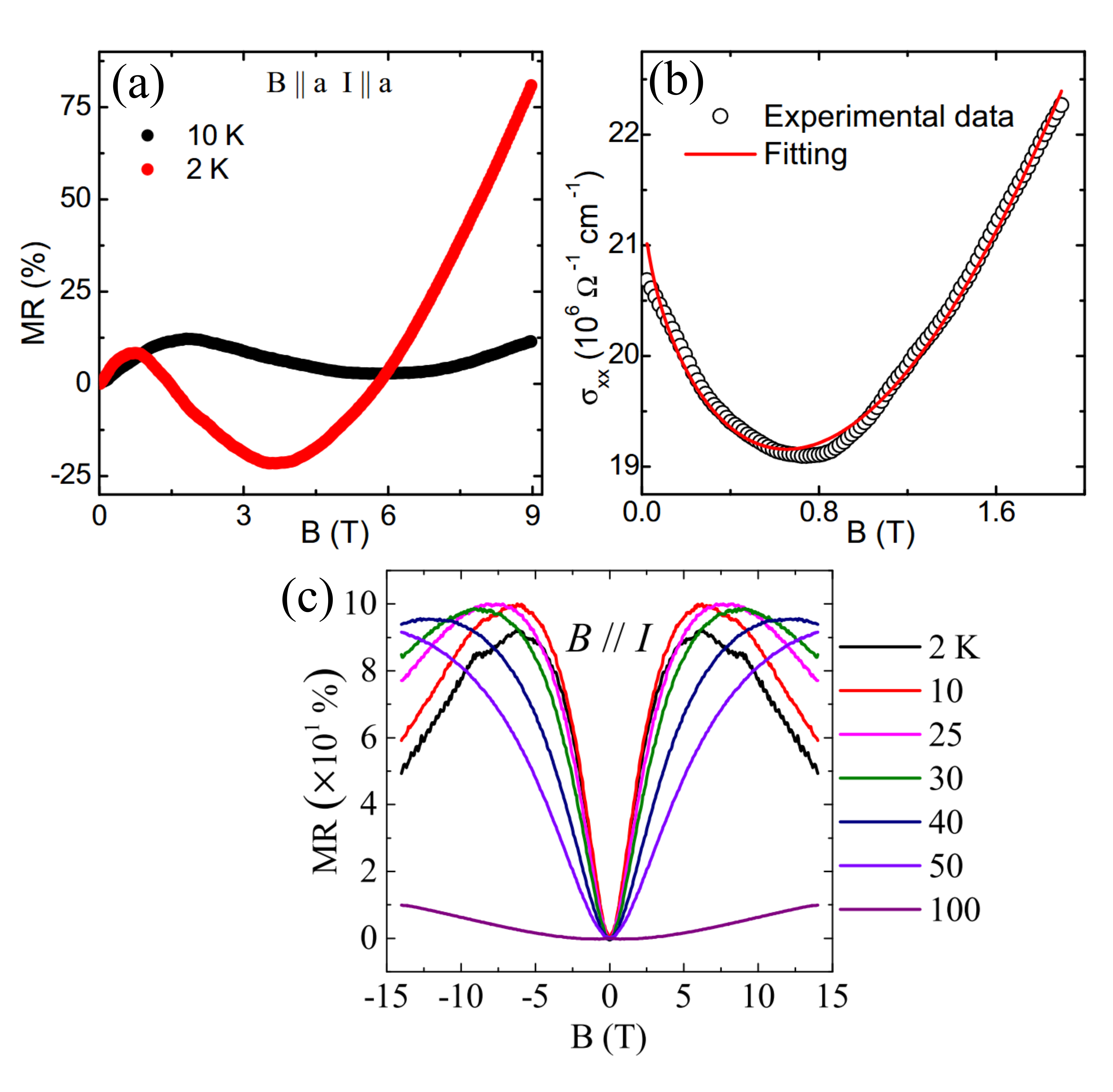}
\caption{Macroscopic manifestation of chiral anomaly
(a) Longitudinal MR with current (I) and magnetic field (B) applied parallel along the $a$ axis at temperatures $T=2$ and $10$ K in ZrSiS, (b) Fitting of the longitudinal magnetoconductivity at 2\, K using the semiclassical formula of chiral anomaly \cite{son2012chiral} and (c) The observed negative magnetoresistance (MR) in ZrAs$_2$ at various temperatures has been attributed to the chiral anomaly \cite{ mi2025extremely}. This figure is adapted from reference \cite{singha2017large, mi2025extremely}, and the relevant permissions have been obtained for this review article.
}
\label{fig11} 
\end{figure}

The application of parallel electric and magnetic fields breaks the conservation of chirality, resulting in an imbalance between opposite chiralities\cite{son2012chiral}. In the realm of condensed matter physics, topological semimetals provide the solid-state realization of this phenomenon \cite{xiong2015evidence,armitage2018weyl}. These materials exhibit linear and inverted band structures that intersect at isolated points or along loops, thereby characterizing different classes of topological semimetals, including Dirac, Weyl, and nodal-line semimetals. In a Weyl semimetal, pairs of Weyl nodes appear in the bulk band structure. Under a magnetic field, the energy bands split into Landau levels, including a chiral zeroth Landau level at each node. An electric field aligned with the magnetic field drives quasiparticles from one chirality node to counter-chirality nodes, resulting in the pumping of charge carriers between opposite chiralities and creating an imbalance in axial chemical potential. In transport measurements, such an imbalance leads to negative longitudinal magnetoresistance (NLMR) and is viewed as the macroscopic manifestation of the chiral anomaly \cite{hirschberger2016chiral,tanwar2023gravitational}. Recently, NLMR has been observed in nodal-line semimetals and attributed to the chiral anomaly as reported in FIG.~\ref{fig11} \cite{singha2017large, mi2025extremely}. It is worth noting that NLMR should be interpreted with caution, since it can also arise from extrinsic artifacts, particularly in ultrahigh-mobility systems where current jetting may mimic a chiral-anomaly-like response \cite{naumann2020orbital}. Therefore, NLMR alone may not provide definitive evidence of an intrinsic chiral anomaly. Complementary and more robust macroscopic probes, such as heat-transport measurements, should also be considered to establish the intrinsic nature of anomaly-related transport signatures.
In this context, a similar effect (positive magneto-thermal conductivity) arises when a thermal gradient replaces an electric field. In this scenario, a mixed axial gravitational anomaly facilitates energy transfer between Weyl cones, resulting in an anomalous heat current \cite{gooth2017experimental, tanwar2023gravitational}. Remarkably, these anomalous responses remain connected through the Wiedemann–Franz law \cite{chernodub2022thermal, tanwar2023gravitational, vu2021thermal}, which describes the relationship between electrical and thermal conductivities that has been recognized for nearly two centuries. Intuitively, such an effect is also expected to occur in nodal-line semimetals. This phenomenon should be investigated for the macroscopic signature of the chiral anomaly in thermal conductivity measurements, which serves as a definitive manifestation of the chiral anomaly. Thus, topological semimetals serve as experimentally tunable platforms for exploring quantum anomaly physics.

\section{Prospective role of Resonant Inelastic X-ray Scattering (RIXS) in probing nodal-line topology}
RIXS is a photon-in and out spectroscopic technique, yielding momentum- and energy-resolved spectra of charge-neutral excitations\cite{de2024resonant}. It allows for polarization-dependent measurements, which reveal how photons with different polarizations interact with matter. This interaction provides valuable information about the nature of the excitations, including phonon, magnon, and d-d type excitations, among others. The unique capabilities of RIXS, along with its element selectivity, make it a complementary technique to neutron-based scattering methods. However, due to its small focus, RIXS requires only a small sample comparable to its focal area.
RIXS provides valuable information on electronic wave functions, especially on spin and orbital selective scattering. For example, the detection of pseudospin structure and the bonding/antibonding nature of quasimolecular\cite{kim2009phase} orbitals in quantum dimers \cite{kim2014excitonic,kim2017resonant}. This technique relies on the principle of coherent interference among various intermediate states \cite{revelli2019resonant}.

Currently, RIXS has been widely used to measure the topological indices of nodal points in topological semimetals \cite{kourtis2016bulk,schuler2023probing}. In this direction, ARPES is a widely used technique for mapping the electronic band structure, predominantly probing the surface rather than the bulk. The topological nodes are buried deep inside the bulk band structure. Therefore, alternative methods need to be explored to visualize the bulk band structure and identify topological features. Before focusing on topological properties, it is important to determine whether bulk band effects in topological semimetals can be detected using RIXS.
The RIXS intensity is expected to be constrained by crystalline symmetry eigenvalues, which determine the band topology. Such constraints lead to distinct scattering intensity for a particular momentum and energy, allowing for the reconstruction of the topological band indices \cite{kourtis2016bulk,lee2023metrology}. Furthermore, the unique spin-orbital selectivity of polarized RIXS allows for the measurement of the effective spinor components close to the nodal point/loop. 

Nodal-line semimetals are promising for RIXS due to their ability to access dynamic aspects of nodal-line physics through momentum- and energy-resolved loss spectra \cite{shen2024resonant}. Key features include particle-hole continua reflecting nodal manifold structure, collective charge modes like plasmons, and momentum-dependent linewidth renormalization of phonons and magnons linked to nodal-line carriers \cite{de2024resonant}.

RIXS is becoming essential for studying charge, spin, and lattice dynamics in quantum materials, with recent studies showing that broad continua can be analyzed via particle-hole scattering in topological semimetals \cite{shen2024resonant}. Evidence for collective excitations, such as nodal-line plasmons in ZrSiS, has been observed \cite{xue2021observation}, aided by advancements in instrumentation for better energy resolution \cite{kim2024advances, schunck2024compact, wang2025high}. Challenges persist, including edge-dependent cross sections, overlapping continua affecting background subtraction, and issues with high-flux operation impacting sample integrity \cite{de2024resonant, shen2024resonant}.

\begin{figure}[h]
\centering
\includegraphics[width=0.8\linewidth]{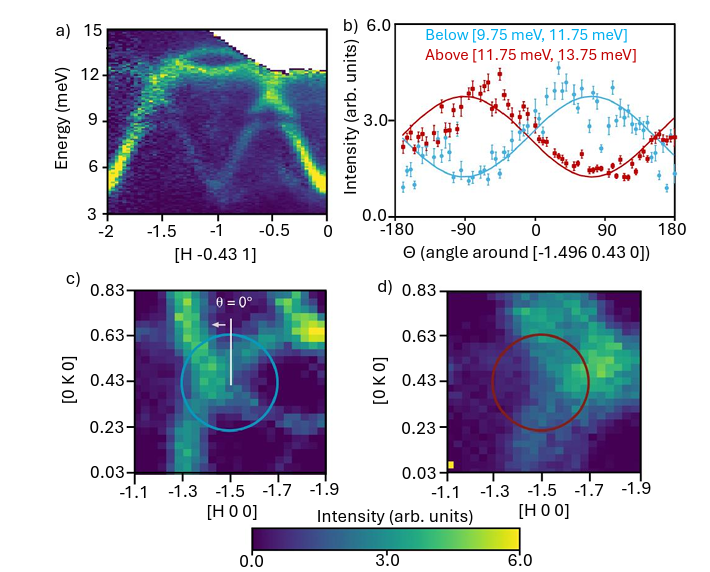}
\caption{(a) Magnon dispersion along the $[H\,{-}0.429\,1]$ direction intersects nodal lines at $(-1.50,\,0.429,\,0)$. (b–d) INS intensity is integrated within a 2\,meV window above (red) and below (blue) the crossing, shown as a function of angle $\alpha$ around the nodal point. Data include one-sigma error bars, with solid curves representing refined sinusoidal fits that exhibit a bimodal pattern. $k_x$-$k_y$ maps at $k_z=0$ (c,d) illustrate energy dependence around the crossing, with circles marking the integration region in (b). The white line in (c) indicates $\alpha=0$ and its increasing direction. Adapted from reference \cite{fahmy2025topological} with relevant permissions.
}
\label{fig12}
\end{figure}

Recently, the concept of nodal-line quasiparticles has been expanded from electronic to magnonic excitations. In this context, the non-coplanar antiferromagnet MnTe$_2$ has been found to exhibit a similar nodal-line topology in its magnonic band structure, as reported in FIG.~\ref{fig12}. 

In this system, low-energy spin excitations create symmetry-protected loops of band degeneracy. The presence of topological magnon nodal lines has been confirmed through inelastic neutron scattering (INS) and magneto-Raman spectroscopy, which are characterized by a pseudo-spin winding in the scattering intensity. Furthermore, when exposed to an external magnetic field, these magnon nodal lines evolve into Weyl magnons, demonstrating the universality of nodal-line physics across both electronic and magnetic quasiparticles. Additionally, phonons with energies between $10$ and $15$ meV along the M–R momentum direction overlap with the magnon branches \cite{fahmy2025topological}. This suggests an interesting avenue for future investigation, where phonon–magnon coupling could drive hybridization between lattice and spin excitations, leading to the formation of magnon–polarons. Such hybridized excitations may further produce measurable macroscopic signatures in heat transport, similar to those observed in Weyl semimetals \cite{tanwar2025emergent}.

In the broader context of future research, RIXS would complement INS and enable precise detection of magnonic nodal lines in MnTe$_2$ through momentum-resolved energy-loss features across the entire Brillouin zone, rather than just near the $\Gamma$-point. This technique directly couples to magnetic excitations at the Mn L-edge, clearly revealing nodal-line crossings as loop-shaped dispersions and intensity modulations, thereby supporting insights from INS. Unlike INS, RIXS offers element-specific sensitivity with micron-sized spot, allowing access to higher-energy excitations and more detailed mapping of topological magnon band structure. Thus, RIXS  would provide better visualization of magnonic phenomena, complementing the validation offered by INS.

\section{Current Status of Nodal-Line Semimetal Materials}

In this section, we summarize the representative examples of nodal-line semimetals, classifying them into experimentally established materials, SOC-tunable or partially gapped realizations, and theoretically proposed or emerging candidates. A limited number of compounds was firmly established experimentally, where ARPES and complementary experiments provide consistent evidence of nodal-line band crossings. These include ZrSiS \cite{Neupane16}, ZrSiSe, ZrSiTe \cite{hosen2017tunability} ZrSnTe \cite{ZrSnTe19}, and HfSiS \cite{Takane16}, Chalcogenide (NiSe) \cite{Ali2016SciAdv}, as well as dipnictides such as ZrAs$_2$ \cite{Wadge21}, ZrP$_2$ \cite{bannies2021extremely}, HfAs$_2$ \cite{muhammad2024electronic} and HfP$_2$ \cite{sims2020termination}.The observation of Fermi-level Kramers nodal lines in rhombohedral TMDCs represents an important recent extension of experimentally accessible nodal-line materials~\cite{Domaine2025}. Other experimentally supported cases include mirror-symmetry-protected systems such as PbTaSe$_2$ \cite{Bian16a}, PbTaS$_2$ \cite{Zhao24}, Mg$_3$Bi$_2$ \cite{Chang19}, two dimensional single layer ferromagnet GdAg$_{2}$ \cite{PhysRevLett.123.116401} the two-dimensional Cu$_2$Si \cite{Feng17}. Magnetic and strongly correlated systems further enrich the classification, with experimentally studied examples such as elemental Co \cite{clark2026manifold}, YMn$_2$Ge$_2$ \cite{yang2024topological}, and the XSbTe family \cite{Regmi23,Regmi24,Elius25,elius2025electronic,valadez2025low} demonstrating that nodal-line features can persist or evolve under magnetic ordering and spin-orbit coupling.

Among these materials, some exhibit nodal-line structures sensitive to SOC, leading to partial gap openings. These include heavier ZrSiX members such as ZrSiTe \cite{hosen2017tunability}, dipnictides such as HfP$_2$ \cite{sims2020termination} and ZrP$_2$ \cite{bannies2021extremely}, and rare-earth antimonide tellurides (XSbTe family), where nodal features evolve continuously across the series but are not strictly gapless.

A large set of theoretically proposed or partially explored systems includes Ca$_3$P$_2$ \cite{Xie15}, CaP$_3$ \cite{Xu17}, and Cu$_3$(Pd,Zn)N \cite{Kim15}, TlTaSe$_2$ \cite{Bian16a}, Eu$_{5}$Bi$_{3}$\cite{wu2019weyl} nonsymmorphic and double-nodal-line systems such as SrIrO$_3$ \cite{Fang15,Chen2015,PhysRevB.97.081105}, BaMX$_3$ compounds \cite{Liang16}, and Zr$_5$Pt$_3$ \cite{Bhattacharyya2021-qy} and hourglass or exotic dispersions in materials like X$_3$SiTe$_6$ \cite{Li18XSiTe}, Ag$_{2}$BiO$_{3}$ \cite{PhysRevB.98.075146} and ReO$_2$ \cite{Hirai23}. In addition to spectroscopic investigations, several magnetic nodal-line semimetals, including  Mn$_3$GaC  \cite{gangwar2025evidences}, MnAlGe \cite{guin20212d}, Fe$_3$Ge \cite{li2025giant}, and Fe$_3$CuTe$_2$ \cite{Kim2018NatMater}, have experimentally demonstrated anomalous Hall effects through transport measurements, highlighting the strong interplay between topology, Berry curvature, and magnetic order in these systems.

Recent studies have revealed that nodal-line band crossings can be a feature of altermagnetic systems. Depending on the underlying symmetry and spin structure, these materials can host spin-valley-locked nodal lines \cite{hu2026observationspinvalleylockednodal}, symmetry-protected nodal loops \cite{Antonenko25,Qu25Altermagnetic,Fernandes24}, and extended nodal networks \cite{Qu25Altermagnetic}, as demonstrated in both first-principles predictions \cite{He24Z3} and recent ARPES experiments \cite{hu2026observationspinvalleylockednodal}.

\section*{Summary and outlook}

Nodal-line semimetals represent a distinctive class of topological quantum materials in which band crossings extend along one-dimensional manifolds in momentum space. In contrast to Dirac and Weyl semimetals, where topological degeneracies occur at isolated points, nodal-line systems host extended band-touching structures such as loops, lines, chains, or more complex networks whose stability is governed by crystalline and internal symmetries. As discussed throughout this review, these symmetry protections, including mirror, glide, screw, and other nonsymmorphic operations, generate a rich landscape of topological phases characterized by unique bulk invariants and unconventional surface states. A central feature of nodal-line semimetals is the emergence of drumhead-like surface states that arise from the projection of nodal loops onto the surface Brillouin zone. These states are associated with enhanced surface density of states and may provide a fertile environment for correlation-driven phenomena such as magnetism, superconductivity, or electronic reconstruction. At the same time, the torus-like Fermi surfaces and nontrivial Berry phases inherent to nodal-line systems produce characteristic signatures in quantum oscillations, magnetotransport, and anomalous response functions. These macroscopic manifestations provide important experimental routes for probing topological band structures beyond direct spectroscopic imaging. Experimentally, the development of high-resolution ARPES has played a decisive role in establishing nodal-line semimetals as a concrete materials platform. Systematic mapping of the three-dimensional electronic structure has enabled the visualization of extended band crossings and the identification of symmetry-protected degeneracies across multiple families of compounds, including square-net materials, transition-metal dipnictides, and rare-earth antimonide tellurides.
In many cases, the combination of ARPES with magnetotransport measurements has provided a coherent picture linking microscopic band topology with measurable electronic properties. Another key theme emerging from current research is the tunability of nodal-line topology. Spin–orbit coupling, lattice distortions, chemical substitution, and external pressure can modify or gap nodal manifolds, transforming nodal-line semimetals into related topological phases such as Dirac, Weyl, or topological insulating states. Magnetic ordering introduces an additional layer of complexity by breaking time-reversal symmetry and generating new classes of magnetic topological semimetals. These mechanisms illustrate how symmetry reduction can act as a powerful control parameter for engineering electronic topology and exploring topological phase transitions (see summary table \ref{tab:NLSM_overview}).
\begin{table*}[t]
\centering
\scriptsize
\setlength{\tabcolsep}{1.6pt}
\renewcommand{\arraystretch}{1}

\caption{Integrated overview of representative nodal-line semimetals (NLSMs), summarizing symmetry protection, spin--orbit coupling (SOC) response, electronic character, key experimental evidence (including transport), and current status. Reference numbers correspond to the citations in the main text.}
\label{tab:NLSM_overview}

\begin{tabular}{l l l l l l l}
\hline\hline

Material & Symmetry & SOC/NLs & Character & Key Evidence & Signatures & Status / Ref. \\

\hline

ZrSiX & Nonsymmorphic & Small gap & Non-mag. & theory, &$\pi$ Berry phase, SdH, & Established  \\
(X=S,Se,Te) & (glide/screw) & (10--20 meV) & & ARPES, transport& large MR, AHE, & ~\cite{schoop2016dirac,hosen2017tunability, fu2019dirac, Neupane16} \\
 &  &  & & &  chiral anomaly,  &  \\
 &  &  & & &  Diamond FS, dHvA  &  \\[6pt]

MPn$_2$ & Nonsymmorphic & Mixed & Non-mag. & theory, & Multiple NLs, $k_z$ & established  \\
 (M=Zr, Hf,& (glide/screw) & (robust + gap) &  &ARPES, transport  & dispersion, vHs, SdH &  \cite{Wadge21,hossain2025superconductivity, bannies2021extremely, sims2020termination, muhammad2024electronic}\\
 Pn=P, As) &  &  &  & & chiral anomaly, SEBC & \\[6pt]

RSbTe (R= & Glide + $P\tilde{T}$ & Tunable & Magnetic & theory, & SOC evolution, & Emerging  \\
Pr, Nd, Gd,& symmetry & (10--60 meV) &  & ARPES, transport& magnetic NLs &\cite{Regmi24,yuan2024observation, Regmi23, hosen2018discovery, Elius25, valadez2025low, elius2025electronic}\\
Tb, Dy, Er)&  &  &  &  &  & \\[6pt]

YMn$_2$Ge$_2$ & $P\tilde{T}$ + glide & Robust & AFM & theory, ARPES & 4-fold Dirac NLs & Established \cite{yang2024topological} \\[6pt]

SmMn$_2$Ge$_2$ / & Magnetic + & Likely & AFM & Limited ARPES & Predicted NL & \\
TbMn$_2$Ge$_2$ & nonsymmorphic & robust &  & (developing) & evolution & not developed \\[6pt]

Elemental Co & Mirror  & Robust & FM & theory, & Spin-polarized NLs & Established  \cite{clark2026manifold} \\
 & & & & Spin-ARPES, transport & & \\[6pt]
 
CaAgX / & Mirror & Destroyed & Non-mag. & Theory, ARPES & NL $\rightarrow$ Weyl & Emerging \cite{Yamakage16,PhysRevX.11.031017} \\
(X=P, As) & & by SOC &  & & transition & \\[6pt]

SrIrO$_3$ & Nonsymmorphic & Robust & Non-mag. & Theory, ARPES & Double NLs & Emerging \cite{Fang15,Chen2015,PhysRevB.97.081105} \\
 & + inversion & & & & & \\[6pt]

MnTe$_2$ &Nonsymmorphic &Robust &AFM &Theory, INS &Weyl magnon pairs & Emerging  \cite{fahmy2025topological} \\
&(glide/screw) &&&Spin-ARPES &plaid-like spin splitting &\\
&inversion &&& & &\\[6pt]
\hline\hline
\end{tabular}
\vspace{1mm}

\footnotesize{
SdH = Shubnikov--de Haas;
MR = magnetoresistance;
AHE = anomalous Hall effect;
FS = Fermi surface;
SEBC = symmetry-enforced band crossings;
SOC = spin--orbit coupling;
dHvA = de Haas--van Alphen;
vHs = van Hove singularities;
NLs = nodal lines.
AFM = antiferromagnetic
FM = ferromagnetic
}
\end{table*}

Despite significant progress, several open questions remain. One ongoing challenge is the unambiguous experimental reconstruction of nodal-line topology throughout the full three-dimensional Brillouin zone, particularly in materials with strong spin--orbit coupling or complex magnetic structures. While surface-sensitive probes such as ARPES have played a central role, complementary bulk-sensitive techniques, including quantum oscillations, optical spectroscopy, and resonant inelastic x-ray scattering, are expected to provide critical insight into dynamical and collective aspects of nodal-line systems. Beyond the characterization of static band structures, nodal-line semimetals offer promising opportunities to explore interaction-driven phenomena. In particular, flat or weakly dispersive electronic states arising from mechanisms such as drumhead surface states, interface reconstruction, moiré superlattices, or proximity to van Hove singularities can strongly enhance the density of states and thereby promote superconducting instabilities, as proposed in early theoretical works on flat-band superconductivity \cite{Khodel1990, Volovik1994, Heikkila2016}. Recent ARPES measurements have directly revealed topological flat bands in rhombohedral graphite \cite{Zhang2024PNAS}, establishing a concrete experimental platform where topology, correlations, and flat-band physics coexist. Furthermore, superconductivity emerging from flat bands has been experimentally demonstrated in moiré systems such as twisted bilayer graphene \cite{Cao2018}, providing a clear proof of principle for flat-band--enhanced pairing.

At the same time, several studies on graphite-based materials have reported intriguing signatures of superconductivity at or near room temperature under ambient pressure \cite{Esquinazi2014, Arnold2018, Kopelevich2024}. However, these observations are often spatially localized, sample-dependent, and not yet fully reproducible, and therefore remain the subject of ongoing debate. As a result, the connection between flat-band physics and high-temperature superconductivity in these systems, while promising, is not yet firmly established. Hybrid approaches that combine flat-band electronic structures with high-energy phonon modes, for example, in hydrogen-rich or interfacial systems \cite{Kawashima2024}, offer an additional pathway toward enhancing pairing scales. More broadly, nodal-line semimetals provide a fertile platform for investigating the interplay between symmetry, topology, and many-body physics. Continued advances in materials design, high-resolution spectroscopy, and theoretical modeling will be essential for clarifying the role of flat-band physics in superconductivity and for identifying robust, reproducible pathways toward high-temperature superconductivity under ambient conditions.

For some compounds, further clarification is required to determine whether they should be classified as altermagnetic or conventional antiferromagnetic systems. Moreover, additional studies on `altermagnetic nodal-line materials' will be valuable for advancing the understanding of their electronic structure and symmetry-related topological properties.
\section*{Acknowledgments}
	
The authors thank to A. Wiśniewski and G. Volovik for insightful comments and suggestions on the manuscript. This research was supported by the “MagTop” project (FENG.02.01-IP.05-0028/23) carried out within the “International Research Agendas” programme of the Foundation for Polish Science co-financed by the European Union under the European Funds for Smart Economy 2021-2027 (FENG). A.S.W acknowledges the support of the National Science Centre, Poland (NCN), through the MINIATURA 9 with project No. 2025/09/X/ST3/00809. A.S.W. also acknowledges support from the National Synchrotron Radiation Centre SOLARIS, which is supported by the Ministry of Science and Higher Education, Poland, under Contract No. 1/SOL/2021/2. C.A. and G.C. acknowledge support from PNRR MUR project PE0000023-NQSTI. 

\section*{Competing financial interests:}

The authors declare no competing financial interests.

\section*{Data availability}
		
No new data were created or analyzed in this review; all information is from previously published studies with proper citations and relevant permissions. Additional materials can be requested from the corresponding authors.
	

\section*{References}
	
        \bibliographystyle{naturemag}
	\bibliography{biblio}
	
	

	\clearpage
	\newpage

	\onecolumngrid

	\setcounter{equation}{0}
	\renewcommand{\theequation}{S\arabic{equation}}
	\setcounter{figure}{0}
    \renewcommand{\figurename}{}
	\renewcommand{\thefigure}{Supplementary Figure~\arabic{figure}}
	\setcounter{section}{0}
	\renewcommand{\thesection}{~\arabic{section}}
	\setcounter{table}{0}
	\renewcommand{\thetable}{S\arabic{table}}
	\setcounter{page}{1}

\end{document}